\documentclass{aastex631}

\usepackage{amsmath}
\usepackage{amssymb}
\usepackage{rotating}
\usepackage{mathtools}
\usepackage{comment}

\begin{document}

\title{A New Framework for Multi-Line Analysis Combined Kernel PCA and Kernel SHAP: 

A Case of NGC 1068 ALMA Band~3 Data}

\author[0009-0007-2898-5365]{Hiroma~Okubo}
\affiliation{Division of Physics, Faculty of Pure and Applied Sciences, University of Tsukuba, Tsukuba, Ibaraki 305-8571, Japan}

\collaboration{20}{}

\author[0000-0001-8416-7673]{Tsutomu~T.~Takeuchi}
\affiliation{Division of Particle and Astrophysical Science, Nagoya University, Furo-cho, Chikusa-ku, Nagoya 464-8602, Japan}
\affiliation{The Research Center for Statistical Machine Learning, The Institute of Statistical Mathematics, 10-3 Midori-cho, Tachikawa, Tokyo 190-8562, Japan}

\author[0000-0002-4623-2718]{Shotaro~Akaho}
\affiliation{National Institute of Advanced Industrial Science and Technology, Tsukuba, 305-8568, Japan}
\affiliation{The Institute of Statistical Mathematics, Tachikawa, 190-8562, Japan}

\author[0000-0002-2501-9328]{Toshiki~Saito}
\affiliation{Faculty of Global Interdisciplinary Science and Innovation, Shizuoka University, 836 Ohya, Suruga-ku, Shizuoka 422-8529, Japan}

\author[0000-0003-1042-6657]{Yasuhiko~Igarashi}
\affiliation{Institute of Engineering, Information and Systems, University of Tsukuba, 1-1-1 Tennodai, Tsukuba 305-8573, Japan}
\affiliation{Tsukuba Institute for Advanced Research (TIAR), University of Tsukuba, 1-1-1 Tennodai Tsukuba, Ibaraki 305-8577, Japan}

\author[0000-0002-1234-8229]{Nario~Kuno}
\affiliation{Division of Physics, Faculty of Pure and Applied Sciences, University of Tsukuba, Tsukuba, Ibaraki 305-8571, Japan}
\affiliation{Tomonaga Center for the History of the Universe, University of Tsukuba, 1-1-1 Tennodai, Tsukuba, Ibaraki 305-8571, Japan}

\author[0000-0002-6824-6627]{Nanase~Harada}
\affiliation{National Astronomical Observatory of Japan, 2-21-1 Osawa, Mitaka, Tokyo 181-8588, Japan}
\affiliation{Astronomical Science Program, Graduate Institute for Advanced Studies, SOKENDAI, 2-21-1 Osawa, Mitaka, Tokyo 181-1855, Japan}

\author[0000-0002-9695-6183]{Akio~Taniguchi}
\affiliation{Kitami Institute of Technology, 165 Koen-cho, Kitami, Hokkaido 090-8507, Japan}

\author[0000-0001-6788-7230]{Shuro~Takano}
\affiliation{Department of Physics, General Studies, College of Engineering, Nihon University, Tamura-machi, Koriyama, Fukushima 963-8642, Japan}

\author[0000-0002-8467-5691]{Taku~Nakajima}
\affiliation{Faculty of Engineering, Suwa University of Science, 5000-1 Toyohira, Chino, Nagano 391-0292, Japan}



\begin{abstract}
We present a new framework for multi-line analysis that combines kernel principal component analysis (Kernel PCA), an unsupervised machine-learning method, and Kernel SHapley Additive exPlanations (Kernel SHAP), an explainable artificial intelligence (XAI) technique.
To enable a comparison with PCA-based studies, which have been widely used in multi-line analyses, we apply our framework to integrated intensity maps of 13 molecular lines from Atacama Large Millimeter/submillimeter Array (ALMA) Band~3 archival data of the nearby galaxy NGC~1068.
Previous PCA-based studies of NGC~1068 reported that physically meaningful structures are mainly captured up to the second component.
In contrast, our framework can interpret physically meaningful features up to the fourth component.
Furthermore, by comparing the results obtained from our framework with molecular column densities derived from local thermodynamical equilibrium (LTE) analysis, we suggest that the abundance of HCO$^+$ is relatively enhanced in the molecular outflow region extending to a radius of about 400~pc from the galactic center, likely due to the effects of ultraviolet radiation and highly dense gas.
These results show that our framework can provide data-driven insights into physical and chemical features that have not been clearly identified in previous studies.
It also provides an efficient tool for interpreting the rapidly increasing amount of multi-line observational data.
\end{abstract}



\section{Introduction} \label{sec:intro}
In recent years, observations with the Atacama Large Millimeter/submillimeter Array (ALMA) have enabled multi-line and high-resolution studies of molecular lines \citep[e.g.,][]{Martin+2021, Izumi+2023}. Molecular line intensities are sensitive not only to the physical conditions of the gas, such as density and temperature, but also to chemical conditions, including variations in chemical composition and molecular reaction networks \citep[e.g.,][]{Penaloza+2018, Harada+2019}. Therefore, molecular line observations provide a powerful tool for investigating how galactic environments are affected by phenomena such as Active Galactic Nucleus (AGN) or star formation feedback \citep[e.g.,][]{Izumi+2013, Butterworth+2025}. Furthermore, studies have been advanced using molecular line ratios to constrain physical and chemical states, or to derive physical quantities under certain assumptions, such as the large velocity gradient (LVG) approximation \citep[e.g.,][]{van+2007, Harada+2024_ratio}.  

A commonly used approach is to take ratios of integrated intensity maps of molecular lines.
By constructing multiple ratio maps and combining them with other observational data, it is possible to infer physical processes occurring within galaxies \citep[e.g.,][]{Garcia+2010, Saito+2022_CI, Stuber+2025}.
However, line ratios depend on multiple parameters, such as gas density, temperature and molecular abundance, and therefore their interpretation is not necessarily unique.
As a result, their interpretation can vary significantly depending on the local environment and position within a galaxy.
Moreover, recent and ongoing high sensitivity, wide bandwidth, and high spatial resolution line surveys with ALMA are rapidly increasing the amount of available information.
These factors make it difficult to apply ratio-based analyses uniformly over large regions or across many molecular line maps.
Therefore, a unified analysis method that can provide a consistent interpretation across wide areas and a large number of molecular line maps is required.

In recent years, feature extraction methods, such as principal component analysis (PCA) and non-negative matrix factorization (NMF), have made it possible to analyze many molecular lines simultaneously \citep[e.g.,][]{Meier+2005, Costagliola+2011, Saito+2022, Harada+2024_pca, Kishikawa+2025, Okubo+2025}. These methods help us to create new maps that contain more information.
Nevertheless, these methods are inherently limited by their assumption of linearity in the relationships among molecular line intensities.
Therefore, in image analysis using PCA, the maps with the first and second components are usually only thought to have physical meaning \citep[e.g.,][]{Costagliola+2011, Okubo+2025}.
However, residual maps with less information often contain several physical processes mixed together. By studying these maps carefully, it may be possible to find complex phenomena that were missed in previous analyses \cite[]{Okubo+2025}.
Therefore, it is important to understand maps with less information, but at present, there is no method that can extract meaningful features from maps with less information.

One reason why it is difficult to extract physical processes correctly is that many complex phenomena occur inside galaxies, causing the data structure to become non-linear.
In contrast, PCA is a linear feature extraction method.
Therefore, in this study, we propose a new analysis method that combines kernel principal component analysis (Kernel PCA), an unsupervised machine learning method, and Kernel SHapley Additive exPlanations (Kernel SHAP), a method of explainable artificial intelligence (XAI), to solve this problem.
Kernel PCA is one of the non-linear feature extraction methods, so it is able to overcome the weakness of PCA \cite[]{Scholkopf+1996}. 
Despite this advantage, Kernel PCA has its own limitation. It cannot tell us which molecular lines strongly contribute to the maps it outputs. 
To solve this, we introduce Kernel SHAP \cite[]{Lundberg+2017}. Kernel SHAP is one of the XAI methods that can explain the “black box’’ of machine learning, and it may help us understand which molecular lines contribute to the maps produced by Kernel PCA.

We apply this framework to the integrated intensity maps of 13 molecular lines from ALMA Band~3 archival data of the nearby galaxy NGC~1068. 
We demonstrate that the combination of Kernel PCA and Kernel SHAP enables the extraction of non-linear features that are not captured by traditional PCA.
Moreover, this new framework allows these features to be physically interpreted.

The structure of this paper is as follows. In Section~\ref{sec:data}, we present the data processing. In Section~\ref{sec:methods}, we describe the analysis methods, Kernel PCA and Kernel SHAP. In Section~\ref{sec:results}, we show the results of our framework. Finally, in Section~\ref{sec:discussion}, we discuss the astrophysical interpretation and implications.

\section{Data Processing}\label{sec:data}

\begin{figure*}[t]

    \begin{center}
        \includegraphics[width=18cm]{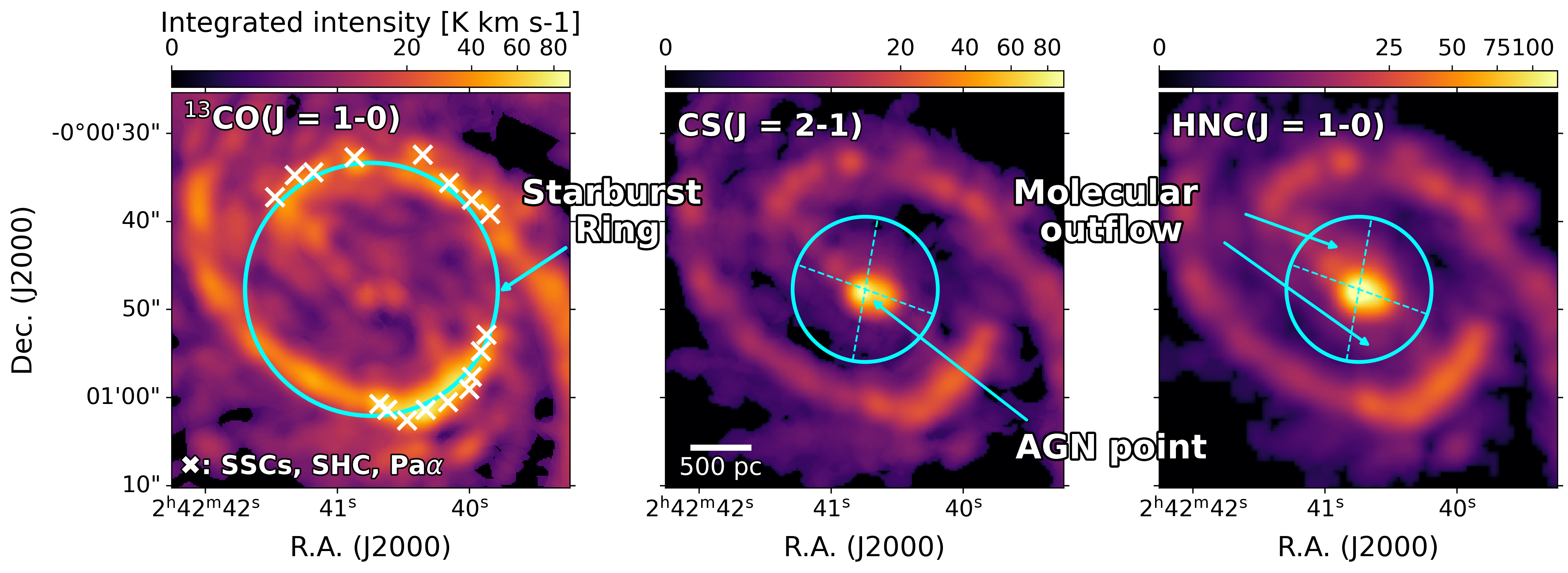}\\
    \end{center}

    \caption{
    The integrated intensity maps of NGC~1068.  
    The left panel presents the $^{13}$CO($J$=1--0) integrated intensity map.  
    The crosses indicate the positions of the SSC, SHC, and Pa$\alpha$ regions defined in \cite{Rico-villas+2021}.  
    Note that the circle marking the starburst ring is not a strict definition, but only shows its approximate location.  
    The middle and right panels show the CS($J$=2--1) and HNC($J$=1--0) integrated intensity maps, respectively.  
    The central circle indicates the field of view (FoV) of the [CI] data.  
    The region outlined with dotted arrows marks the direction of the molecular outflow \citep{Saito+2022_CI, Saito+2022}. The diameter of the circle is 1 kpc.
    }\label{NGC1068_overview}
\end{figure*} 
\begin{figure*}[t]

    \begin{center}
        \includegraphics[width=18cm]{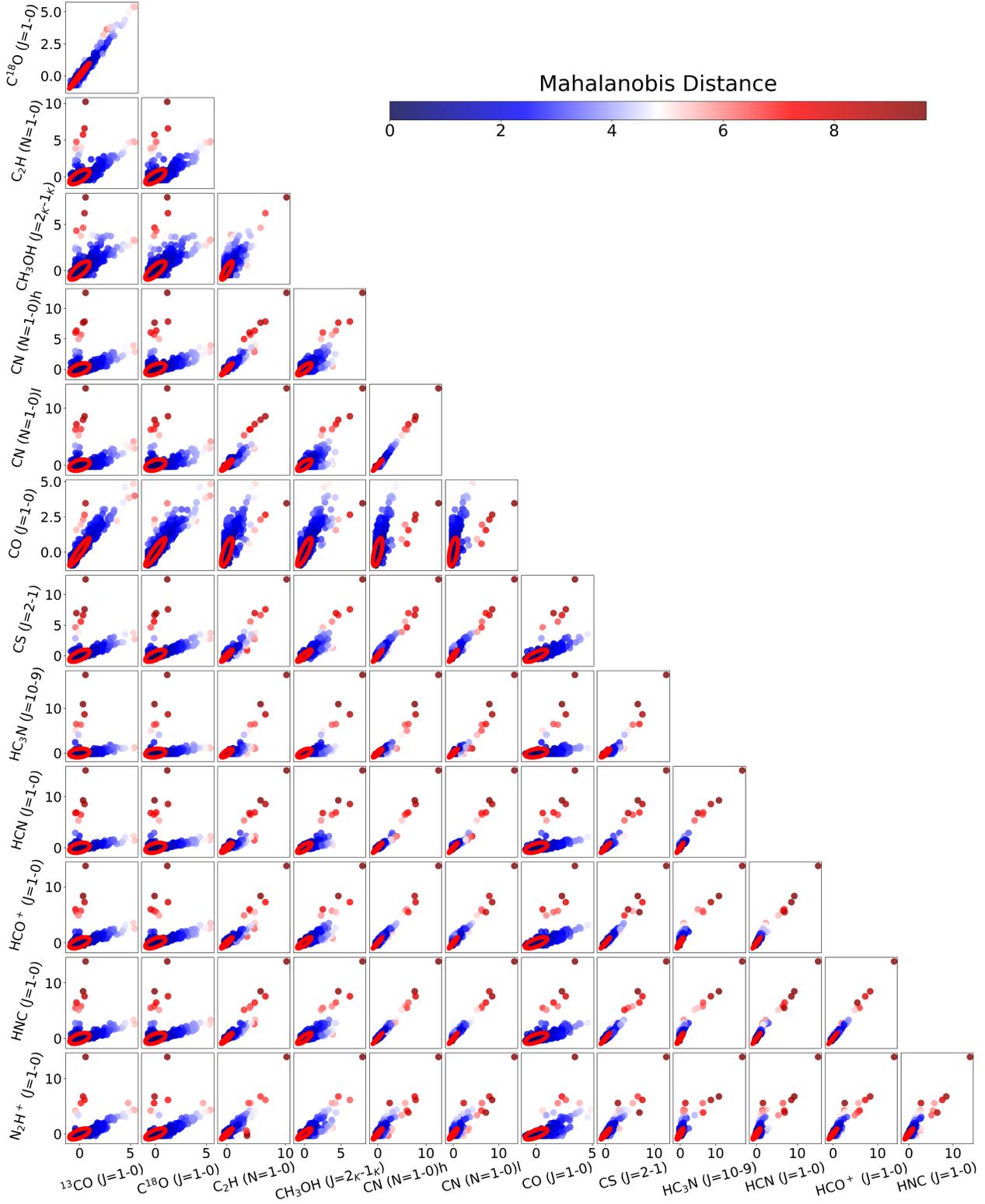}\\
    \end{center}

    \caption{Correlation plot between two standardized molecular lines. The color indicates the Mahalanobis distance, which accounts for the correlation between the two variables. The red ellipse represents a Mahalanobis distance of 1.
    These plots correspond to the maps shown in Figure~1 (left), Figure~2, and Figure~3 of \citet{Okubo+2025}.
    }\label{two_plots}
\end{figure*}
We describe how the data are arranged in this work.
To apply Kernel PCA, it is necessary to define samples and features.
A sample represents a single observational target, while a feature describes a property used to characterize the sample.
In this study, we define each pixel as a sample and the integrated intensities of individual molecular lines as features.
Although this method can also be applied to data cubes, caution is required when the number of features greatly exceeds the number of samples, as this may lead to statistically unstable results.
\footnote{If one wishes to analyze data where the number of features exceeds the number of samples, it is necessary to adopt the framework of high-dimensional statistical analysis (see, e.g., \citealt{Takeuchi+2024}).} 

To test the validity of our new method, we need to compare its results with those obtained from the conventional PCA.  
For this reason, it is desirable to apply our analysis to a dataset that has already been studied with PCA.  
In addition, to show that our method can extract useful information from complex environments, we select a galaxy where multiple physical processes coexist, such as an AGN, molecular outflow, and starburst regions.

Based on these considerations, we select NGC~1068 as our target.
Figure~\ref{NGC1068_overview} shows the ALMA observations of $^{13}$CO($J$=1--0), CS($J$=2--1), and HNC($J$=1--0).  
This galaxy has a starburst ring, where active star formation is taking place.  
Within this ring, superstar clusters (SSCs) and superhot core (SHC) have been found (Figure~\ref{NGC1068_overview}, left) \citep{Rico-villas+2021}.  
At the galactic center, there is AGN activity, and a circumnuclear disk (CND) is located around the AGN (Figure~\ref{NGC1068_overview}, middle).  
In addition, a molecular outflow has also been detected (Figure~\ref{NGC1068_overview}, right).  
In this way, the observed molecular line properties of NGC~1068 suggest that the dominant physical and chemical processes vary from region to region.
Furthermore, the apparent complexity of the molecular line properties can depend on the beam size, implying that careful attention is required when interpreting the results because of beam averaging over multiple physical components. 
In this study, we use the 13 molecular lines: C$_2$H($N$=1--0), HCN($J$=1--0), HCO$^+$($J$=1--0), HNC($J$=1--0), HC$_3$N($J$=10--9), N$_2$H$^+$($J$=1--0), CH$_3$OH($J$=2$_{\it K}$--1$_{\it K}$), CS($J$=2--1), C$^{18}$O($J$=1--0), $^{13}$CO($J$=1--0), CN($N_J$=1$_{3/2}$--0$_{1/2}$), CN($N_J$=1$_{1/2}$--0$_{1/2}$), and CO($J$=1--0).
For simplicity, hereafter we denote the $N_J$=1$_{3/2}$--0$_{1/2}$ and $N_J$=1$_{1/2}$--0$_{1/2}$ transitions as $N$=1--0h and $N$=1-0l, respectively.
When the h or l notation is not given, the fine-structure lines are treated collectively.

The data were reduced using the Physics at High Angular resolution in Nearby GalaxieS (PHANGS)--ALMA pipeline. 
We first regridded the maps to a spatial scale of 150~pc, so that one grid cell approximately corresponds to one to a few giant molecular clouds (GMCs).
This regridding helps reduce spatial correlations between neighboring grid cells, allowing us to treat each cell as an independent sample. 
After this step, we standardized all molecular-line maps to equalize the variance among the features. 
We also note that, for simplicity, we use one-dimensional Kernel PCA in this study.
For more details on the data reduction, we refer the readers to \citet{Leroy+2021} and \citet{Saito+2022}.

This final dataset forms a matrix of size 683 × 13, where 683 is the number of samples (grids) and 13 is the number of features (molecular lines).
The dataset used in this study is identical to that constructed in Section~2.1 of \citet{Okubo+2025}.
The standardized integrated intensity maps from \citet{Okubo+2025} are shown in Figures~1 (left), 2, and 3, and their statistical properties are summarized in Table~1.

\section{Methods}\label{sec:methods}
We first summarize the limitations of applying conventional PCA to multi-line analysis in Section~\ref{subsec:Limitation of conventional PCA}.  
In Section~\ref{subsec:Kernel PCA}, we introduce Kernel PCA as a method that can address these limitations of conventional PCA, and discuss both its effectiveness and the remaining challenges when applied to multi-line data.  
Finally, in Section~\ref{subsec:Kernel SHAP}, we introduce Kernel SHAP to overcome these challenges and explain how Kernel PCA and Kernel SHAP, which are independent methods, can be combined in this study.

We define a common dataset, denoted by $\mathbf{X}$, which is used as the input for both PCA and Kernel PCA.
\begin{equation}
\mathbf{X} = [\mathbf{x}_1,\ldots,\mathbf{x}_N] \in \mathbb{R}^{D\times N},
\label{eq:data_matrix}
\end{equation}
As described in Section~\ref{sec:data}, the dataset $\mathbf{X}$ is composed of standardized integrated intensities of ALMA Band~3 molecular lines in NGC~1068 ALMA Band~3 for each spatial grid.
The vector $\mathbf{x}_n$ represents the standardized integrated intensities of each molecular line at the $n$-th grid.
Unless otherwise specified, all vectors are treated as column vectors.
The number of features $D$ corresponds to the number of molecular lines and is set to $D = 13$ in this study.  
The number of samples $N$ corresponds to the number of spatial grids and is $N = 683$.

Note that we use the \texttt{scikit-learn} package (version 1.5.2; \citealt{Pedregosa+2011}) and the \texttt{shap} library (version 0.46.0; \citealt{Lundberg+2017}) to perform Kernel PCA and Kernel SHAP, respectively.
\subsection{Limitation of conventional PCA}\label{subsec:Limitation of conventional PCA}
\subsubsection{Basic Theory}
We briefly summarize the theoretical basis of PCA and then discuss its intrinsic limitations.
We define the covariance matrix $\mathbf{C}$ of the dataset $\mathbf{X}$ as
\begin{equation}
\mathbf{C} = \frac{1}{N}\mathbf{X}\mathbf{X}^{\mathrm T}
\in \mathbb{R}^{D\times D}.
\label{covariance matrix}
\end{equation}
PCA is formulated as an eigenvalue problem of the covariance matrix $\mathbf{C}$,
\begin{equation}
\mathbf{C}\,\boldsymbol{\omega}_i = \lambda_i \boldsymbol{\omega}_i,
\label{eigenvalue problem}
\end{equation}
where $\lambda_i$ and $\boldsymbol{\omega}_i \in \mathbb{R}^{D}$ are the $i$-th eigenvalue and its corresponding eigenvector, respectively.
In the following, the $i$-th principal-component score (hereafter PC$i$ score) of the n-th grid, $y_i(\mathbf{x}_n)$, is defined as
\begin{equation}
y_i(\mathbf{x}_n)
=
\boldsymbol{\omega}_i^{\mathrm{T}} \mathbf{x}_n.
\label{eq:pca_score}
\end{equation}

This quantity corresponds to the grid values of the PCA-extracted map (hereafter PC$i$ map).
For each PC$i$ map, the overall contribution of individual molecular lines can be quantified by the eigenvector $\boldsymbol{\omega}_i$, which represent the relative importance of each line.

For example, the PC$i$ maps shown in Figure~7 of \citet{Okubo+2025} visualize the PC$i$ scores defined in Eq.~(\ref{eq:pca_score}).
These maps extract characteristic regions by combining information from multiple molecular lines.
In addition, the contribution of each molecular line to a given PC$i$ map can be understood by examining the corresponding eigenvectors shown in Figure~8 of \citet{Okubo+2025}.
Assuming linear relationships among the integrated intensities of molecular lines, the PC$i$ maps and their corresponding eigenvectors can be used
to discuss which molecular lines dominate each PC$i$ map and what physical and chemical features are extracted.

\subsubsection{Mahalanobis distance} \label{subsubsec:Mahalanobis distance}
The Mahalanobis distance is a distance measure that takes into account the variance and correlations of the data.
For a data vector $\mathbf{x}$ with mean $\boldsymbol{\mu}$ and covariance matrix $\mathbf{C}$,
the Mahalanobis distance $d_{\mathrm{M}}$ for the $n$-th data sample is defined as
\begin{equation}
d_{\mathrm{M},n}
=
\sqrt{
(\mathbf{x}_n - \boldsymbol{\mu})^\mathrm{T}
\mathbf{C}^{-1}
(\mathbf{x}_n - \boldsymbol{\mu})
}.
\end{equation}
This distance gives smaller values along directions with large variance and larger values along directions with small variance.
Therefore, it can be interpreted as a statistical distance that reflects the shape of the data distribution.

Figure~\ref{two_plots} shows two-dimensional scatter plots of the 13 standardized integrated intensities used in this study.
In each panel, the red ellipse represents a Mahalanobis distance of $d_M = 1$ calculated within that two-dimensional space.
When PCA is applied to these two-dimensional distributions, the PC1 axis is defined along the major axis of the red ellipse, while the PC2 axis is defined along the minor axis.
In Section~\ref{subsubsec:limitation}, we explain the limitations of applying PCA to multi-line analyses with the Mahalanobis distance.

\subsubsection{Limitation} \label{subsubsec:limitation}
As an illustrative example, we consider the standardized integrated intensities of $^{13}$CO($J$=1--0) and C$^{18}$O($J$=1--0) (Figure~\ref{two_plots}, top left) to explain the geometric meaning of PCA.
If PCA were applied to this distribution, the PC$i$ axes would be aligned with the major and minor axes of the red ellipse ($d_{\mathrm{M}} = 1$), as described in Section~\ref{subsubsec:Mahalanobis distance}.
This illustrates that, when the variance structure of the data is dominated by a single linear correlation between two standardized integrated intensities, PCA is expected to capture the dominant variance effectively.

In contrast, we consider the case of PCA applied to the standardized integrated intensities of $^{13}$CO($J$=1-0) and N$_2$H$^+$($J$=1-0) (Figure~\ref{two_plots}, bottom left).  
In this panel, this dataset exhibits two distinct linear trends, characterized by a shallow-slope trend and a steep-slope trend.  
When PCA is applied to such a dataset, the PC1 axis is aligned with the correlation direction of the shallow-slope trend, while the PC2 axis represents the variance perpendicular to this correlation.  
As a result, the steep-slope trend cannot be accurately captured by the PCA axes.

As illustrated in Figure~\ref{two_plots}, many pairs of standardized integrated intensity maps exhibit nonlinear structures arising from the presence of two distinct linear trends.
Such nonlinear structures are likely caused by spatially localized variations in physical parameters, such as temperature, density, and optical depth \citep[e.g.,][]{Pety+2017}.
These results show that, while PCA can capture global trends in integrated intensity maps, it is insufficient to describe detailed relationships among molecular lines.
To capture these complex relationships, nonlinear feature extraction methods are required.

Indeed, \citet{Stuber+2023} reported the presence of two distinct linear trends in the distribution of two high-density gas tracers, HCN and N$_2$H$^+$ in M51.  
They interpreted these trends as reflecting different underlying physical and chemical conditions.  
This result highlights that properly characterizing nonlinear relationships associated with two distinct linear trends is directly linked to an accurate understanding of the physical and chemical states of the GMCs.

\subsection{Kernel PCA}\label{subsec:Kernel PCA}
In Section~\ref{subsec:Limitation of conventional PCA}, we described the theory of conventional PCA and pointed out that PCA has limitations in describing the physical and chemical states of galaxies. We also showed that nonlinear feature extraction methods are required to obtain a more detailed understanding.  
In this section, we explain Kernel PCA, which is the core of our newly proposed framework, by comparing it with PCA.

As described at the beginning of Section~\ref{sec:methods}, we again consider a feature space in which the bases are the standardized integrated intensities of $D$ molecular lines, and $N$ grids are distributed as samples in this space.

\subsubsection{Basic idea}
Kernel PCA is a nonlinear feature extraction method proposed by \citet{Scholkopf+1996}.
The basic idea of Kernel PCA is to map samples into a high-dimensional space using a function $\phi$,
\begin{equation}
\phi: \mathbb{R}^D \rightarrow \mathcal{H},
\label{eq:phi_map}
\end{equation}
and then apply linear PCA in that space.
By mapping the data into a high-dimensional space, nonlinear relationships among samples in the original feature space become easier to treat in a linear way.
It is important to note that the basis defined in this high-dimensional space does not directly correspond to individual molecular lines.

As shown in Eqs.~\eqref{covariance matrix} and \eqref{eigenvalue problem}, PCA requires the computation of inner products for the dataset $\mathbf{X}$.
However, computations in high-dimensional spaces are very expensive, making the direct calculation of inner products impractical.
\footnote{\sloppy
This phenomenon is referred to as the curse of dimensionality.
The curse of dimensionality refers to the phenomenon in which the computational cost increases exponentially with increasing dimensionality, making problems computationally intractable \citep[see e.g.,][]{Bellman1957, Donoho2000}.}
Therefore, in Kernel PCA, the inner product in the high-dimensional space is computed directly using a function in the original feature space,
\begin{equation}
\phi(\mathbf{x}_i)^\top \phi(\mathbf{x}_j)
= k(\mathbf{x}_i, \mathbf{x}_j),
\label{eq:kernel_trick}
\end{equation}
which is known as the kernel trick.
This approach enables nonlinear analysis without explicitly handling the high-dimensional mapping.

\subsubsection{Selection of a kernel function}
The function defined in Eq.~\eqref{eq:kernel_trick} is called a kernel function.
A kernel function can be chosen arbitrarily as long as it satisfies the required conditions \citep{Shawe-Taylor+2004}.
One of the most commonly used kernel functions is the radial basis function (RBF) kernel,
which is defined as
\begin{equation}
k(\mathbf{x}_i,\mathbf{x}_j)
= \exp\!\left( -\gamma \|\mathbf{x}_i-\mathbf{x}_j\|^2 \right),
\label{eq:rbf_kernel}
\end{equation}
where $\gamma$ is the kernel width.
Here, if $\gamma$ is chosen too small, Kernel PCA asymptotically approaches
linear PCA, whereas excessively large values of $\gamma$ lead to increased
sensitivity to noise. Therefore, the choice of $\gamma$ requires careful
consideration.

One of the advantages of the RBF kernel is that it defines the similarity between samples in a high-dimensional space based on the distance between samples in the original feature space.
When samples are close to each other in the original feature space ($\mathbf{x}_i \simeq \mathbf{x}_j$), the inner product approaches unity. The samples are regarded as similar also in the high-dimensional space.
In contrast, when the distance between samples is large, the inner product approaches zero. The samples are regarded as dissimilar.

As described in Section~\ref{subsubsec:limitation}, Figure~\ref{two_plots} shows two distinct linear trends among molecular lines, which are thought to reflect different physical and chemical conditions.
This implies that the distance between samples reflects how different their physical and chemical states are.
For these reasons, we consider that the RBF kernel can appropriately capture the properties of molecular line data. 
Therefore we adopt the RBF kernel as the kernel function in this study.
We note that the aim of this study is not to compare different kernel functions, and a systematic comparison is beyond the scope of this paper.

When the RBF kernel is used, the high-dimensional space becomes infinite dimensional.
This can be understood from the Taylor expansion of the RBF kernel,
\begin{equation}
k(\mathbf{x}_i,\mathbf{x}_j)
\propto
\exp\!\left(2\gamma\,\mathbf{x}_i^\top\mathbf{x}_j\right)
=
\sum_{n=0}^{\infty}
\frac{(2\gamma)^n}{n!}
\left(\mathbf{x}_i^\top\mathbf{x}_j\right)^n .
\end{equation}

\subsubsection{Comparison with PCA}
The matrix constructed by a kernel function is called the Gram matrix $\mathbf{K}$.
The $(i,j)$ element of the Gram matrix $\mathbf{K}_{ij}$ is defined as
\begin{equation}
\mathbf{K}_{ij} = \phi(\mathbf{x}_i)^\top \phi(\mathbf{x}_j)
= k(\mathbf{x}_i, \mathbf{x}_j).
\label{eq:kernel_matrix_def}
\end{equation}
In Kernel PCA, we consider deviations from the mean in the high-dimensional space.
Therefore, the Gram matrix is centered as
\begin{equation}
\tilde{\mathbf{K}}
= \mathbf{K}
- \mathbf{1}_N \mathbf{K}
- \mathbf{K} \mathbf{1}_N
+ \mathbf{1}_N \mathbf{K} \mathbf{1}_N
\label{eq:centered_kernel}
\in \mathbb{R}^{N\times N}.
\end{equation}
Here, $\mathbf{1}_N$ is an $N \times N$ matrix whose elements are all equal to $1/N$.
Kernel PCA is performed by solving the eigenvalue problem of the centered Gram matrix $\tilde{\mathbf{K}}$,
\begin{equation}
\tilde{\mathbf{K}}\,\boldsymbol{\mathbf{a}}_i
= \lambda_i N \boldsymbol{\mathbf{a}}_i,
\label{eq:eigen_problem}
\end{equation}
where $\lambda_i$ and $\boldsymbol{\mathbf{a}}_i$ are the eigenvalue and eigenvector, respectively.
Note that $\boldsymbol{\mathbf{a}}_i$ is the eigenvector of the centered Gram matrix defined in the sample space, rather than in the original feature space.

While PCA solves an eigenvalue problem of the covariance matrix $\mathbf{C}$,
Kernel PCA solves an eigenvalue problem of the centered Gram matrix $\tilde{\mathbf{K}}$.
This is the essential difference between PCA and Kernel PCA.

In PCA, the PC$i$ score is defined by Eq.~(\ref{eq:pca_score}).
In Kernel PCA, the $i$-th kernel principal component score (hereafter KPC$i$ score) is defined as
\begin{equation}
y_i(\mathbf{x}_n)
= \sum_{j=1}^N a_{ij}\, k(\mathbf{x}_n,\mathbf{x}_j).
\label{eq:kpca_score}
\end{equation}
We also refer to maps constructed from the KPC$i$ scores as the KPC$i$ maps.

\subsubsection{Indicator of information content}
The contribution ratio (CR) and the cumulative contribution ratio (CCR) are metrics used to quantitatively evaluate the extracted features obtained from PCA and Kernel PCA.
These metrics measure how much information contained in the original dataset $\mathbf{X}$ is preserved.

The eigenvalue $\lambda_i$ represents the variance explained by component $i$.
CR is
\begin{equation}
\mathrm{CR}_i
= \frac{\lambda_i}{\sum_{j=1}^{N} \lambda_j},
\label{eq:cr}
\end{equation}
and CCR is
\begin{equation}
\mathrm{CCR}(m)
= \sum_{i=1}^{m} \mathrm{CR}_i.
\label{eq:ccr}
\end{equation}

Here, the CR and CCR in PCA are defined based on the variance of samples in the original feature space, and therefore reflect the variation in molecular line intensities.
In contrast, in Kernel PCA, these quantities are based on the variance of samples in a feature space mapped to high dimensions by a kernel function, and should be interpreted not as physical variances but as indicators of the concentration of information.

\subsubsection{Critical issue}
As shown above, Kernel PCA can capture nonlinear relationships among the integrated intensities of molecular lines.
However, Kernel PCA has an important limitation.
It is difficult to identify which molecular lines contribute to the extracted KPC$i$ maps.

In conventional PCA, the basis of the feature space is described by the standardized integrated intensities of individual molecular lines \citep[e.g.,][]{Harada+2024_pca, Okubo+2025}.
Therefore, by inspecting the eigenvectors $\boldsymbol{\omega}$, it is possible to understand the overall contribution of each molecular line to the PCA maps.
In contrast, in Kernel PCA, the basis in the high-dimensional space does not correspond directly to the standardized integrated intensities of individual molecular lines.
Furthermore, when the RBF kernel is used, the high-dimensional space becomes infinite-dimensional.
In this case, the basis vectors cannot be computed explicitly.
As a result, the contribution of individual molecular lines to the KPC$i$ maps cannot be directly evaluated.

This is a critical issue for multi-line analysis.
Multi-line analysis relies on the physical and chemical properties of individual molecular lines to interpret the internal conditions of galaxies \citep[e.g.,][]{Harada+2019, Saito+2022}.

To address this issue, we introduce Kernel SHAP, a XAI method.
The details are described in the next section.

\subsection{Kernel SHAP}\label{subsec:Kernel SHAP}
In Section~\ref{subsec:Kernel PCA}, we described the theoretical background of Kernel PCA and the critical issues in multi-line analysis.
In the new framework proposed in this study, we introduce Kernel SHAP to address this critical issue.

In Section~\ref{subsubsec:Basic theory of Kernel SHAP}, we explain the basic theory of Kernel SHAP,
and in Section~\ref{subsubsec:Combination between Kernel PCA and Kernel SHAP}, we describe how Kernel PCA and Kernel SHAP are combined.
As in the previous sections, we consider a dataset $\mathbf{X}$ with $D$ features and $N$ samples,
but for simplicity, we focus on the $n$-th sample in the following discussion.

\subsubsection{Basic theory of Kernel SHAP} \label{subsubsec:Basic theory of Kernel SHAP}
SHAP is one of the XAI methods that has attracted attention as a way to address the black-box nature of
machine learning models \citep{Lundberg+2017}.
The objective of SHAP is to represent the output of a machine learning model as a linear combination of the contributions from individual features.
These contributions are referred to as SHAP values.
In this study, the machine learning corresponds to Kernel PCA, and the model outputs are the KPC$i$ scores.
The SHAP values represent the contribution of each molecular line for each grid.

However, the original SHAP values are very expensive to compute, and in most cases it is not possible to calculate them precisely. 
Therefore, in this study, we use Kernel SHAP as an approximate method.

\paragraph{(i) General outline.}
First, for $D$ molecular lines, we introduce a $(D+1)$ dimensional mask vector $\mathbf{z}=(z_0,\dots,z_D)$, which represents whether each molecular line is used.
Here, $z_j=1$ means that the $j$-th molecular line is used in the analysis, and $z_j=0$ means that it is not used.
Note that $z_0$ is always set to $1$, since it corresponds to the baseline.
From an astronomical point of view, this corresponds to a virtual operation that reproduces a situation in which a molecular line is not observed.
Based on this mask vector, we construct a pseudo input $h(\mathbf{z})$ by excluding a subset of molecular lines from the original observational data.
That is, $h(\mathbf{z})$ represents a virtual multi-line observational dataset in which specific molecular lines are missing.

Next, we define $f$ as the Kernel PCA model trained in advance using the original dataset $\mathbf{X}$.
We define the KPC$i$ score obtained by applying this model $f$ to the pseudo input $h(\mathbf{z})$ as $f(h(\mathbf{z}))$.
An important point is that the model is not retrained, and only the input data are changed for the evaluation.
This procedure corresponds to examining how the KPC$i$ score changes when some molecular lines are not observed.

The purpose of Kernel SHAP is to approximate these pseudo KPC$i$ scores using a linear approximation model defined on the mask vector, $\mathbf{z} \boldsymbol{\psi}_n$.
Here, $\boldsymbol{\psi}_n$ represents the contribution of each molecular line at the $n$-th grid, that is, the SHAP values.
We emphasize that Kernel SHAP is a locally additive explanation method, and its additive form is introduced only for interpretability, not to imply physically linear relationships among molecular lines.

By repeating the generation of pseudo inputs and the evaluation of KPC$i$ scores randomly $T$ times, we obtain a set of pseudo KPC$i$ scores for various combinations of molecular lines.
The SHAP values $\boldsymbol{\psi}_n$ are then estimated by optimizing a loss function so that the linear approximation model best matches these samples.

Mathematically, loss function is defined as
\begin{equation}
    L(\boldsymbol{\psi}_n)
    =
    ( \mathbf{f} - \mathbf{Z}\boldsymbol{\psi}_n )^{\top}
    \mathbf{W}
    ( \mathbf{f} - \mathbf{Z}\boldsymbol{\psi}_n ).
    \label{eq:loss function}
\end{equation}
Here, the vectors and matrices are defined as
\begin{equation}
\mathbf{f} =
\begin{pmatrix}
    f(h(\mathbf{z}^{(1)})) \\
    f(h(\mathbf{z}^{(2)})) \\
    \vdots \\
    f(h(\mathbf{z}^{(T)}))
\end{pmatrix}
\in \mathbb{R}^{T},
\quad
\mathbf{Z} =
\begin{pmatrix}
    \mathbf{z}^{(1)} \\
    \mathbf{z}^{(2)} \\
    \vdots \\
    \mathbf{z}^{(T)}
\end{pmatrix}
\in \{0,1\}^{T \times (D+1)},
\quad
\mathbf{W} =
\mathrm{diag}\!\left(
 w(\mathbf{z}^{(1)}),\,
 w(\mathbf{z}^{(2)}),\,
 \dots,\,
 w(\mathbf{z}^{(T)})
\right) \in \mathbb{R}^{T \times T}.
\end{equation}
$w(\mathbf{z}^{(t)})$ is the corresponding weight.
This weight is defined as
\begin{equation}
    w(\mathbf{z}^{(t)}) =
    \frac{D - 1}{\binom{D}{|S^{(t)}|}\, |S^{(t)}|\,(D - |S^{(t)}|)},
\label{eq:shapley kernel}
\end{equation}
and is referred to as the Shapley kernel.
Here, $S^{(t)}$ is the set of features used in $\mathbf{z}^{(t)}$, and $|S^{(t)}|$ is its size.
Focusing on the denominator, we find that the weight becomes large when the number of selected features is close to $0$ or $D$.
 $\mathbf{Z}\boldsymbol{\psi}_n$ can also be interpreted as a linear approximation of the KPC$i$ score at the $n$-th grid.

By minimizing the loss function defined in Eq.~\eqref{eq:loss function}, the SHAP values are obtained as
\begin{equation}
    \boldsymbol{\psi}_n
    =
    (\mathbf{Z}^{\top} \mathbf{W} \mathbf{Z})^{-1}
    \mathbf{Z}^{\top} \mathbf{W} \mathbf{f}
    \in \mathbb{R}^{(D+1)}.
    \label{eq:shap value}
\end{equation}
 
 The \texttt{shap} library sets the default number of samples to $T = 2D + 2048$, and we adopt this default setting in this study.

\paragraph{(ii) Practical approximation.}
Next, we describe how the function $h(\mathbf{z})$ is constructed.
In this study, we determine the background values using the k-means method implemented in the \texttt{shap} library.
With this approach, $h(\mathbf{z})$ is defined as
\begin{equation}
h(\mathbf{z}) = [\, \mathbf{x}_{n,S},\ \mathbf{b}_{m(n),\bar{S}} \,].
\label{eq:hz}
\end{equation}
Here, $\mathbf{x}_{n,S}$ represents the set of features with $z=1$, that is, the features selected to be used in the model.
On the other hand, $\mathbf{b}_{m(n),\bar{S}}$ corresponds to the features with $z=0$, which are not used and are replaced by background values.

The background vector $\mathbf{b}_{m(n),\bar{S}}$ is chosen from
\begin{equation}
B = \{\mathbf{b}_1, \mathbf{b}_2, \dots, \mathbf{b}_M\}, \quad
\mathbf{b}_m \in \mathbb{R}^{D},
\end{equation}
where $B$ is the set of cluster centers obtained by applying k-means clustering to the dataset $X$ with $M$ clusters.
The index $m(n)$ denotes the cluster to which the data point $\mathbf{x}_n$ belongs.
Note that the notation in Eq.~(\ref{eq:hz}) is used only for clarity.
In practice, the input vector preserves the original feature order $j = 1, \dots, D$.
For each feature $j$, the original value $\mathbf{x}_{n,j}$ is used if $j \in S$, while the background value $\mathbf{b}_{m(n),j}$ is used if $j \in \bar{S}$, where $\bar{S}$ represents the set of features not included in $S$.
Therefore, we evaluate the model output as
\begin{equation}
f(h(\mathbf{z})) \approx f([\mathbf{x}_{n,S},\ \mathbf{b}_{m(n),\bar{S}}]),
\end{equation}
and the SHAP values are then estimated using Eq.~(\ref{eq:shap value}).

\subsubsection{Combination between Kernel PCA and Kernel SHAP}\label{subsubsec:Combination between Kernel PCA and Kernel SHAP}
So far, we have described the basic theory of Kernel PCA and Kernel SHAP.
In this section, we finally explain how these two independent methods are combined in our analysis.

A trained Kernel PCA model, which provides the KPC$i$ score for an arbitrary input vector $\mathbf{x}$, can be written as
\begin{equation}
    f(\mathbf{x}) \coloneqq y_i(\mathbf{x})
    = \sum_{j=1}^N a_{ij}\, k(\mathbf{x}, \mathbf{x}_j),
    \label{eq:KPCA score model}
\end{equation}
where the eigenvalues $\lambda_i$ and eigenvectors $\boldsymbol{\mathbf{a}}_i$ of Gram matrix $\mathbf{K}$ are already learned from the dataset $\mathbf{X}$.
Therefore, for the input vector constructed using Eq.~(\ref{eq:hz}),
the model output becomes
\begin{equation}
    f(\mathbf{z}^{(t)}) \coloneqq
    \sum_{j=1}^N a_{ij}
    \exp\!\left(
    -\gamma \bigl\|
    [\mathbf{x}_{n,S^{(t)}},\ \mathbf{b}_{m(n),\bar{S}^{(t)}}]
    - \mathbf{x}_j
    \bigr\|^2
    \right).
    \label{eq:combination}
\end{equation}

By substituting Eq.~(\ref{eq:combination}) into Eq.~(\ref{eq:shap value}), we can evaluate the SHAP values, which represent the contribution of each molecular line to the KPC$i$ score at the $n$-th grid.
Furthermore, by applying this procedure to all $N$ grids for a given component $i$, we can quantify the contribution of molecular lines to the entire KPC$i$ map.

As a result, the contribution of individual molecular lines to the KPC$i$ scores, which is difficult to interpret using Kernel PCA alone, can be evaluated by combining Kernel PCA with Kernel SHAP.
This is the main strength of the new framework proposed in this study.

\begin{figure*}[t]

    \begin{center}
        \includegraphics[width=13cm]{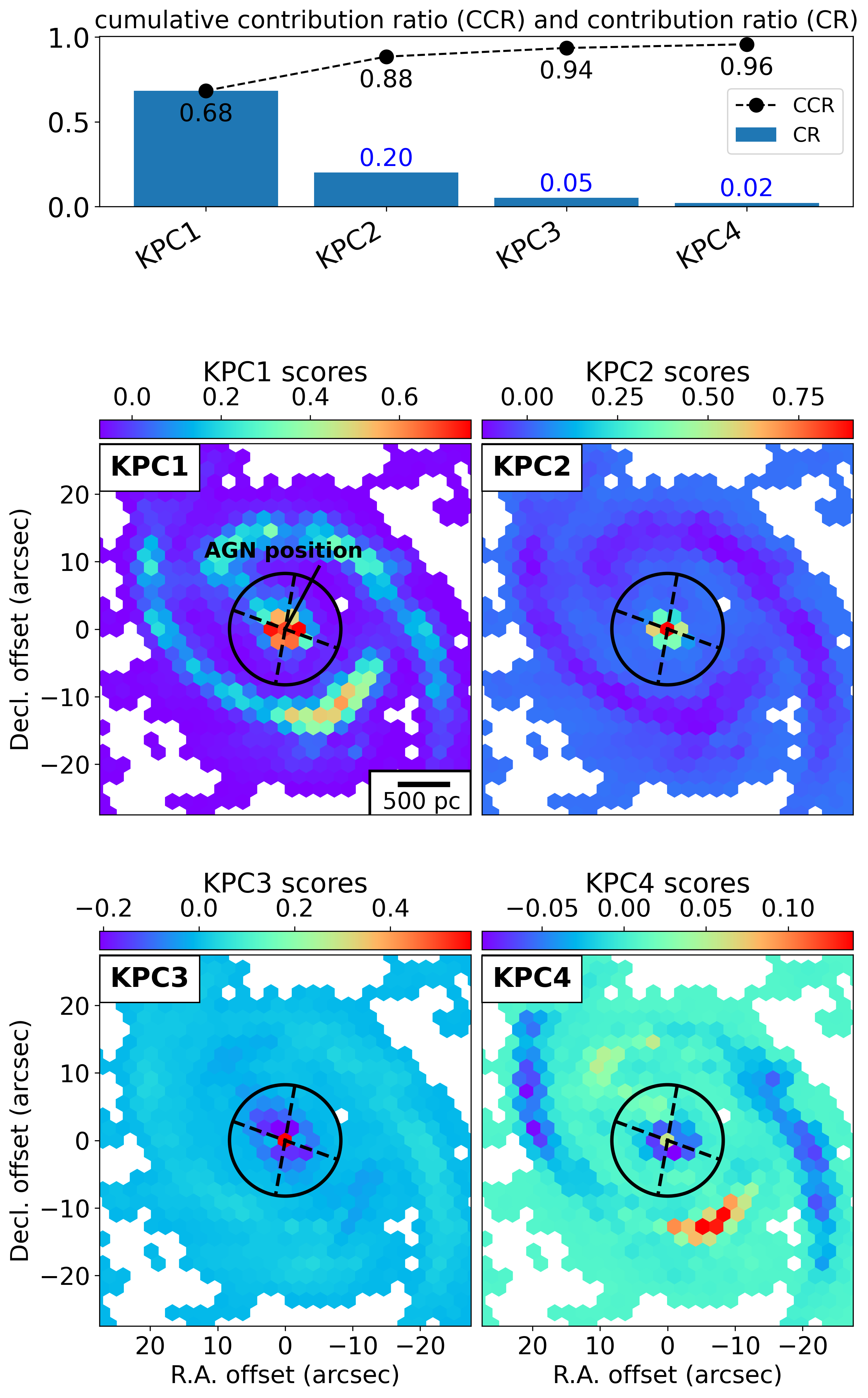}\\
    \end{center}

    \caption{Kernel PCA resutls. The first row shows the contribution ratio (CR) and cumulative contribution ratio (CCR). The second row displays the KPC1 map on the left and the KPC2 map on the right, while the third row shows the KPC3 map on the left and the KPC4 map on the right. The color represents the KPC$i$ scores. The black circle marks the field of view (FoV) of the [CI] data \citep[][]{Saito+2022_CI}. The crossed dashed lines denote the direction of the outflow \citep[][]{Saito+2022}.}\label{KPCA_results}
\end{figure*}

\section{Results} \label{sec:results}

\subsection{Kernel PCA results}\label{subsec:kernel PCA results}
In this work, the kernel width is determined based on astrophysical knowledge.  
\citet{Okubo+2025} applied PCA to the same dataset of this study and showed that PC1 map and PC2 map represent physically meaningful structures, while the PC3 map and the subsequent lower-variance components contain mixed information.
Therefore, we tested several kernel widths in exponential steps ($10^{-5}$, $10^{-4}$, $10^{-3}$, ...), and selected the value for which KPC1 map and KPC2 map reproduce the same spatial patterns as PC1 and PC2 map, while KPC3 map shows a different, new structure compared to PC3 map.  
As a result, we adopt kernel width = $10^{-3}$ for this study.

To evaluate how much information each component preserves from the original dataset, we calculated the CR and the CCR in Eqs.~\eqref{eq:cr} and \eqref{eq:ccr}. 
Since the CCR reaches more than 95\% by KPC4 map, we analyze up to the fourth component in this paper.

The CR, CCR and KPC$i$ maps are shown in Figure~\ref{KPCA_results}.  
The top panel presents the CR and CCR for each KPC$i$ map.  
The second row shows the KPC1 map (left) and KPC2 map (right), and the third row shows the KPC3 map (left) and KPC4 map (right).
Each grid in the KPC$i$ map corresponds to the KPC$i$ score calculated using Eq.~\eqref{eq:kpca_score}.

\begin{figure*}[t]

    \begin{center}
        \includegraphics[width=18cm]{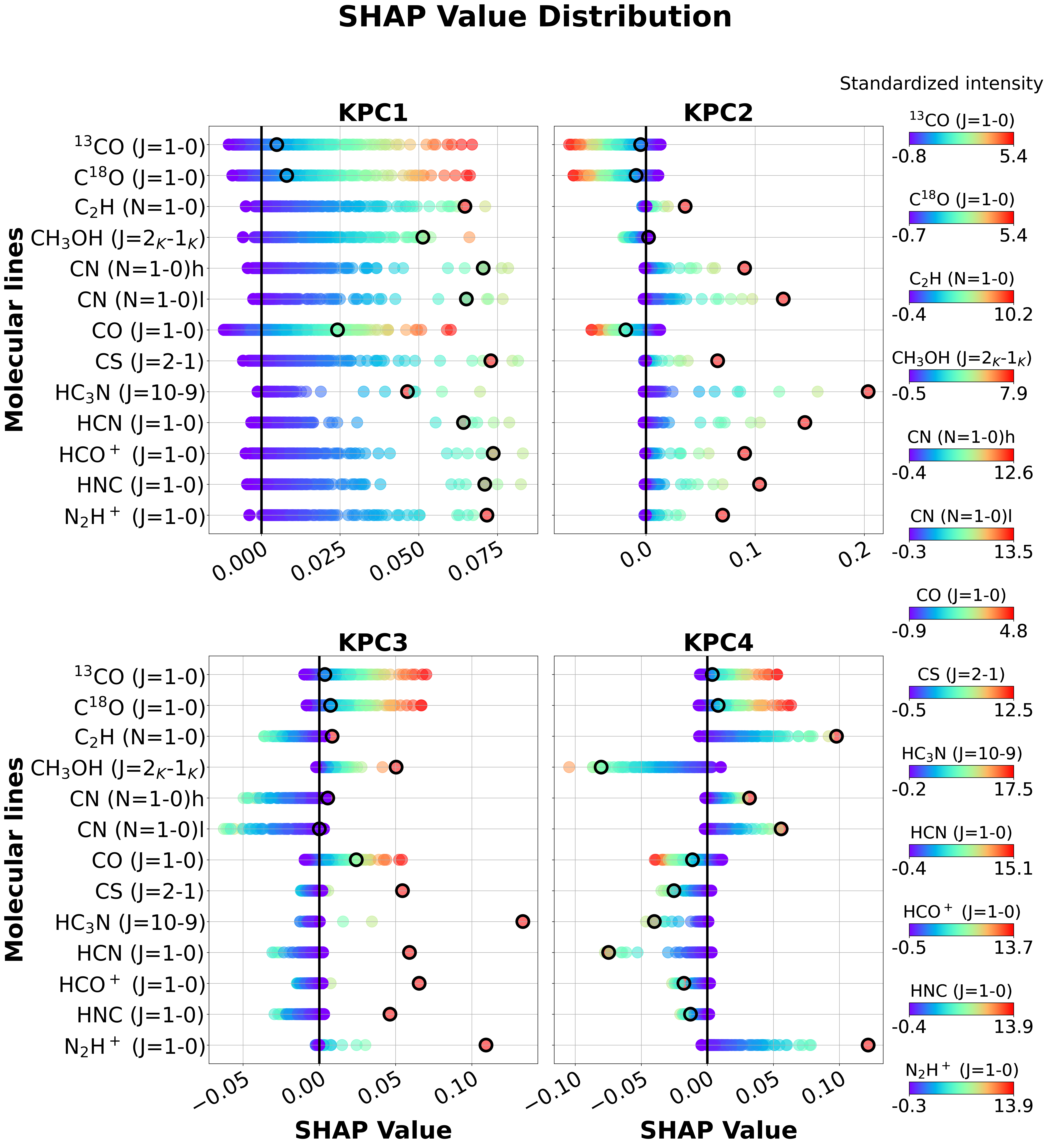}\\
    \end{center}

    \caption{Kernel SHAP resutls. The top row shows the SHAP values for the KPC1 scores on the left and for the KPC2 scores on the right, while the second row shows those for the KPC3 scores on the left and for the KPC4 scores on the right. The color represents the standardized integrated intensity of each molecular line used in the analysis. In each figure, the x-axis corresponds to the SHAP value and the y-axis corresponds to the molecular lines. The black-outlined points indicate the grid at the AGN position.}\label{SHAP value distribution}
\end{figure*}

\paragraph{KPC1.}
The KPC1 map reflects the overall gas distribution of the galaxy, showing positive scores in gas-rich regions.
\paragraph{KPC2.}
The KPC2 map shows high positive scores in the AGN and the CND, while negative scores are mainly distributed along the starburst ring.
\paragraph{KPC3.}
The KPC3 map shows high positive scores around the AGN, lower positive scores near the arms, and negative scores elongated toward the central region.
\paragraph{KPC4.}
The KPC4 map shows strong positive scores at the southwestern bar-ends, while negative scores are enhanced in the CND and arms.

In this study, we do not strictly define the boundaries of the CND, arms, and bar ends.  
Instead, we define the region surrounding the AGN position as the CND,  
the area with KPC2 negative scores as the starburst ring,  
the area with KPC4 positive scores as the bar ends,
and the area with outsides of KPC4 negative scores as the arms.

\subsection{Kernel SHAP results}\label{subsec:kernel SHAP results}
Figure~\ref{SHAP value distribution} shows the SHAP values for the KPC1--KPC4 maps.
The vertical axis represents the molecular lines, while the horizontal axis indicates the SHAP values.
Each sample corresponds to an individual grid, and the color represents the standardized integrated intensity in each grid.
This figure provides an overview of how each molecular line contributes to each KPC$i$ map.
Hereafter, we refer to this figure as the SHAP value distribution.

\paragraph{KPC1.}
In the SHAP value distribution of KPC1 map, all molecular lines show positive values, suggesting that they contribute to the positive scores of the KPC1 map, which correspond to the overall gas distribution of the galaxy.
\paragraph{KPC2.}
For KPC2 map, CO isotopologues and CH$_3$OH($J$=2$_{\it K}$--1$_{\it K}$) exhibit negative SHAP values, indicating their contributions to negative scores of the KPC2 map, corresponding to the starburst ring. 
In contrast, other molecular lines show positive values, associated with the positive scores of the KPC2 map, which correspond to the AGN and the CND.
\paragraph{KPC3.}
In KPC3 map, CN($N$=1--0) and C$_2$H($N$=1--0), followed by HNC($J$=1--0) and HCN($J$=1--0), have negative SHAP values, suggesting their contribution to the negative scores elongated toward the central region. 
Meanwhile, HC$_3$N($J$=10--9), N$_2$H$^+$($J$=1--0), HCO$^+$($J$=1--0), and HCN($J$=1--0) exhibit high positive SHAP values at the AGN position, whereas CO isotopologues show low positive values.
These molecular lines are therefore associated with the positive scores of the KPC3 map, corresponding to the AGN and the arms.
Details on which molecular lines trace the arm and AGN regions are discussed in Section~\ref{subsubsec:KPC3}.
\paragraph{KPC4.}
KPC4 map shows negative SHAP values for CH$_3$OH($J$=2$_{\it K}$--1$_{\it K}$), HCN($J$=1--0) and CO($J$=1--0), implying that these lines contribute to the negative scores of the KPC4 map, corresponding to the CND and the arms. 
In contrast, N$_2$H$^+$($J$=1--0), C$_2$H($N$=1--0), CN($N$=1--0), $^{13}$CO($J$=1--0), and C$^{18}$O($J$=1--0) exhibit positive SHAP values, suggesting their association with the positive scores of the KPC4 map, corresponding to the bar-ends.
Further interpretation of these spatial associations is provided in Section~\ref{subsubsec:KPC4}.

\section{Discussion} \label{sec:discussion}
Although this framework provide a powerful way to interpret some features from the data, they are not derived from the first principles of physics. Therefore, it is crucial to examine whether the these results have physically and chemically meaningful interpretations.

\begin{figure*}[t]

    \begin{center}
        \includegraphics[width=15cm]{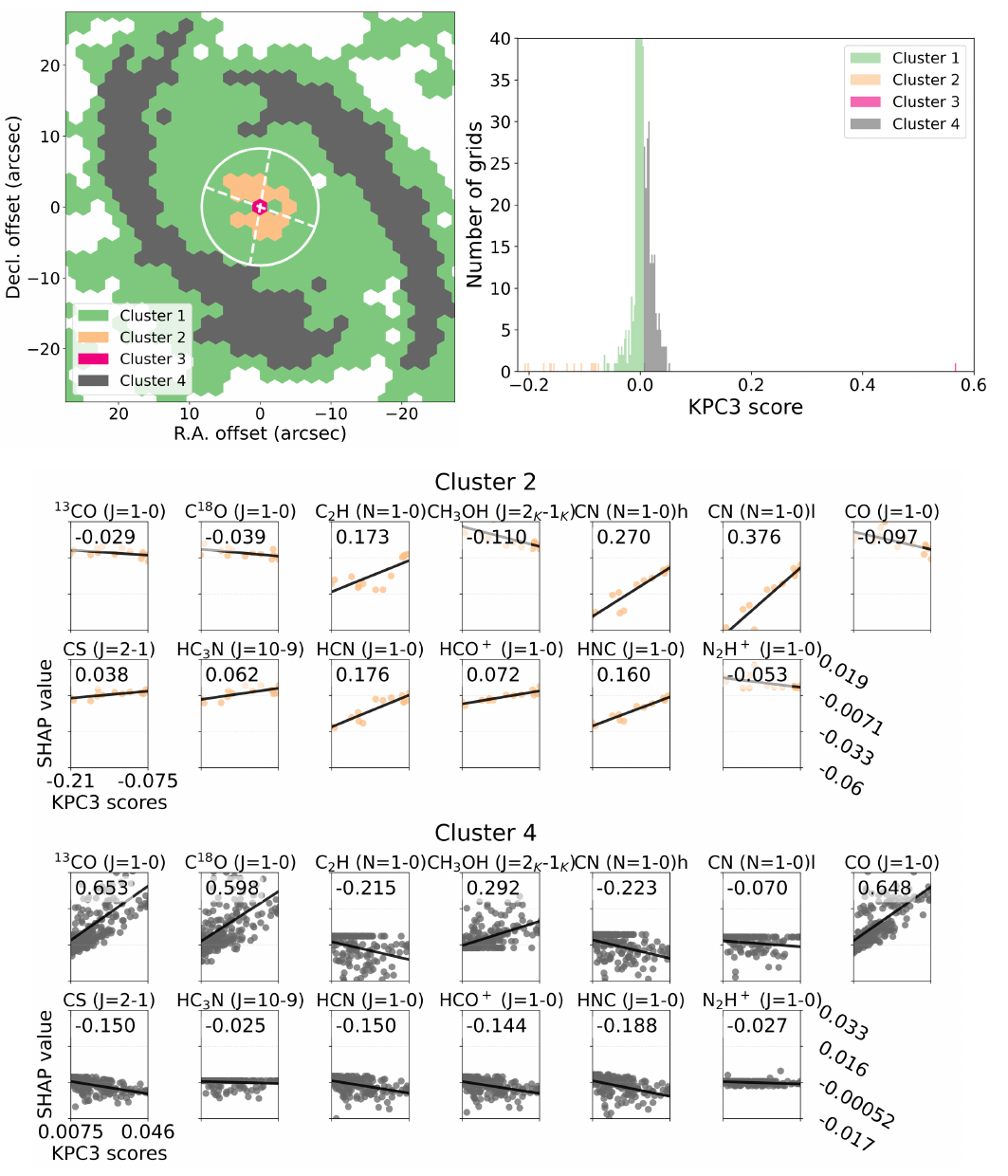}\\
    \end{center}

    \caption{The left panel in the first row shows the cluster map, while the right panel presents the clustering result. The second row illustrates the relationship between the SHAP values and the scores in Cluster~2, and the third row shows the corresponding relationship for Cluster~4. The numbers in the plot represent the correlation coefficients.
    }\label{KPC3_clustering}
\end{figure*}

\subsection{Kernel PCA and Kernel SHAP}\label{subsec:Kernel PCA and Kernel SHAP}

\subsubsection{KPC1 and KPC2}\label{subsubsec:KPC1 and KPC2}
As described in Section~\ref{subsec:kernel PCA results}, we set the kernel width to $10^{-3}$ so that the KPC1 and KPC2 maps correspond well to the PC1 and PC2 maps, respectively.
In the KPC1 map, positive scores are observed in regions with high molecular hydrogen column densities. In addition, the SHAP values of all molecular lines contribute positively in these regions. These results suggest that the KPC1 map primarily reflects information related to the molecular hydrogen column density.
The KPC2 map shows positive scores mainly in the CND and AGN regions, while negative scores correspond to the starburst ring. The Kernel SHAP analysis indicates that CO isotopologues predominantly contribute negative values, whereas highly dense gas tracers such as HC$_3$N($J$=10--9) and HCN($J$=1--0) show positive contributions. Therefore, the KPC2 map is interpreted as mainly reflecting gas density information.
However, in regions such as the CND, molecular abundances can vary in a complex manner due to various chemical processes \citep[e.g.,][]{Izumi+2013, Garcia+2014, Saito+2022}. 
While the KPC2 map primarily reflects density information, abundance-related information may be captured by other component maps (see Section~\ref{subsubsec:KPC4}).

Here, we focus on the SHAP value distributions shown in Figure~\ref{SHAP value distribution},
which represent the contributions of molecular lines to the overall KPC1 and KPC2 scores.
These SHAP value distributions show very good agreement with the PC1 and PC2 coefficients
shown in Figure~8 of \citet{Okubo+2025}.
The signs of the contributions from individual molecular lines are the same in both methods,
and their relative magnitudes also show good agreement.
This result indicates that the SHAP value distributions properly reflect the physical information
contained in the KPC1 and KPC2 maps.
Therefore, although Kernel PCA and Kernel SHAP are independent methods, they provide consistent physical interpretations, demonstrating that this framework works appropriately.

\subsubsection{KPC3}\label{subsubsec:KPC3}
The KPC3 map shows characteristic regions with high positive scores, low positive scores, and negative scores. 
The bottom-left panel of Figure~\ref{SHAP value distribution} shows the overall distribution of SHAP values for KPC3 map, but here we perform a more detailed analysis. Specifically, to investigate the SHAP values for each structural component indicated by KPC3 map, we apply the k-means method to the KPC3 map and examine the SHAP values for each cluster. The results are shown in Figure~\ref{KPC3_clustering}.  
The number of clusters is set to four so that the major structures shown in the KPC3 map---such as the AGN, arms, and the elongated central feature---could be separated in a minimal way.

We note that Cluster~1 does not correspond to a physically distinctive region and is therefore not discussed further in this study.
It should also be noted that the interpretation of SHAP values depends on the sign of the KPC$i$ scores.
For regions with positive KPC$i$ scores, molecular lines with positive SHAP values contribute more strongly to the features represented by the score.
In contrast, for regions with negative KPC$i$ scores, molecular lines with negative SHAP values
play a more important role in explaining the features indicated by the score.

\begin{itemize}
    \item Cluster~2: 
Cluster~2 mainly consists of regions with negative scores.
The spatial distribution of this cluster is consistent with the direction of the molecular outflow.
In addition, since the KPC3 scores of Cluster~2 are negative, molecular lines with negative SHAP values can be interpreted as contributing more strongly to the characteristics of Cluster~2.

The second row of Figure~\ref{KPC3_clustering} shows the SHAP value of the KPC3 scores for Cluster~2, where the vertical axis represents the SHAP value and the horizontal axis represents the KPC3 score, shown separately for each molecular line.
From this figure, it is clear that CN($N$=1--0), C$_2$H($N$=1--0), HNC($J$=1--0),
and HCN($J$=1--0) show noticeably slopes.
In other words, as the KPC3 score becomes more negative, the SHAP values of these lines also become more negative.

This result indicates that the characteristics of Cluster~2 are strongly related to these molecular lines and suggests that the KPC3 score captures a certain systematic trend.
Previous studies have reported that CN, C$_2$H, and HNC are enhanced by UV radiation
in the molecular outflow of NGC~1068 \citep[e.g.,][]{Garcia+2017, Saito+2022}.
Based on these results, Cluster~2 is likely extracting properties characteristic of the molecular outflow.
The slopes seen in the second row of Figure~\ref{KPC3_clustering} may reflect changes in molecular abundances.
A detailed analysis of the abundance variations of HCN and other molecules is presented in Section~\ref{subsec: Are the abundances of HCN, HCO$^+$, and CS enhanced by the molecular outflow?}.

    \item Cluster~3: 
Cluster~3 is dominated by regions with high positive scores.
This cluster is spatially coincident with the AGN position.
Since the KPC3 score of Cluster~3 is positive, molecular lines with positive SHAP value can be interpreted as contributing more strongly to the characteristics of Cluster~3.

In the lower left panel of Figure~\ref{SHAP value distribution}, the black-outlined samples indicate grids at the AGN point.
The SHAP values show that HC$_3$N($J$=10--9) and N$_2$H$^+$($J$=1--0) have particularly high contributions in this cluster.
HC$_3$N is expected to be enhanced in regions with high temperature and high density that are shielded from strong radiation \citep[e.g.,][]{Harada+2013}.
Indeed, \citet{Takano+2014} reported strong HC$_3$N emission toward the CND of NGC~1068.
In addition, N$_2$H$^+$ is thought to be enhanced in environments with strong ionization sources such as Cosmic-Ray Dominated Region (CRDR) \citep[e.g.,][]{Zinchenko+2009,Harada+2015}, and strong N$_2$H$^+$ emission has been detected at the CND of NGC~1068 \citep[e.g.,][]{Aladro+2013}.

Molecular lines with the next highest SHAP values, such as HCN  and HCO$^+$,
are also known to be enhanced in various chemical environments \citep[e.g.,][]{Aalto+2007, Izumi+2013}.
Therefore, we conclude that Cluster~3 reflects the complex chemistry in the AGN region.

    \item Cluster~4: 
Cluster~4 represents the low positive scores and corresponds to the arms.  
Since the KPC3 scores of Cluster~4 are positive, molecular lines with positive SHAP values can be interpreted as contributing more strongly to the characteristics of Cluster~4.

The third row of Figure~\ref{KPC3_clustering} shows the SHAP value of the KPC3 scores for Cluster~4.
The SHAP values indicate that CO isotopologues mainly contribute in this region.  
This feature is similar to the negative scores in the KPC2 map, suggesting that the Cluster~4 also extracts the diffuse gas component in the arm regions.
Since this region has been reported to be rich in CO isotopologues, the Kernel PCA and Kernel SHAP results are consistent with the previous studies \citep[e.g.,][]{Takano+2014, Nakajima+2023}.
\end{itemize}

In the KPC3 map, regions with positive scores are found to contain multiple different components.
Regions with high positive scores (Cluster~3) correspond to the AGN point,
while regions with low positive scores (Cluster~4)
correspond to the spiral arms.
This result indicates that Kernel PCA alone cannot properly separate the multiple components contained in KPC3 map.
However, by combining a clustering analysis and examining the SHAP values for each cluster, these components can be separated more clearly, and the contributions of individual molecular lines can be quantified for each component.
Therefore, while conventional PCA has difficulty interpreting
components beyond the PC3 map, this framework demonstrates that the KPC3 map can also be interpreted in a physically meaningful way.

\subsubsection{KPC4}\label{subsubsec:KPC4}

In the KPC4 map, the negative scores appear mainly in the CND and in the spiral arms. These two regions are known to have very different physical and chemical conditions \citep[e.g.,][]{Viti+2014, Scourfield+2020, Huang+2022, Sanchez+2022}. 
To investigate the contributions of molecular lines to these different regions, we perform a clustering analysis using the k-means method, following the same approach as in Section~\ref{subsubsec:KPC3}.
For the clustering, we use the KPC2 and KPC4 scores as input features to separate between the CND and the spiral arms. To make the KPC4 map play a stronger role in the clustering, we multiply the KPC4 scores by a factor of two before running k-means method. 
The results are shown in Figure~\ref{KPC4_in_detail}. 
The top-left panel shows the clustering result on the sky map, while the top-right panel shows the same result on the KPC2–KPC4 plane. 
The lower panels present the distributions of the scores and the SHAP values for each cluster.
From Figure~\ref{KPC4_in_detail}, the arm region is classified as Cluster~3 and the CND as Cluster~4, which indicates that the clustering works well in a spatial sense. We then examine the SHAP values for each cluster to understand their physical meaning. Since the main structures highlighted by KPC4 are the arms, the CND, and the bar-ends region, we focus on Clusters~1, 3, and 4.

\begin{itemize}
    \item Cluster~1: 
Cluster~1 is located near the bar-ends region.
This region is known to host strong star formation near the bar-ends and to contain SSCs \citep[e.g.,][]{Tsai+2012, Rico-villas+2021, Nagashima+2024}.
Therefore, Cluster~1 is considered to extract regions related to star formation activity at the bar-ends.

Next, we examine the characteristics of Cluster~1 based on the SHAP values.
Since the KPC4 scores of Cluster~1 are positive, molecular lines with positive SHAP values
can be interpreted as contributing more strongly to the features of Cluster~1.
The second row of Figure~\ref{KPC4_in_detail} shows the distribution of SHAP values of the KPC4 scores for Cluster~1.
From this figure, $^{13}$CO($J$=1--0), C$^{18}$O($J$=1--0), C$_2$H($N$=1--0), CN($N$=1--0),
and N$_2$H$^+$($J$=1--0) show positive slopes, while CH$_3$OH($J$=2$_K$--1$_K$) shows a negative slope.
This means that the former molecular lines contribute in the direction that explains the characteristics of Cluster~1, whereas CH$_3$OH works in the opposite direction.
$^{13}$CO and C$^{18}$O have higher effective critical densities than $^{12}$CO and trace relatively dense molecular gas.
C$_2$H and CN are known to be enhanced by UV radiation and are typical PDR tracers \citep[e.g.,][]{Fuente+1993, Pety+2005, Beuther+2008}.
In addition, N$_2$H$^+$ is known to trace high-density gas
in star-forming regions \citep[e.g.,][]{Kauffmann+2017, Tafalla+2021}.
On the other hand, CH$_3$OH is easily photodissociated by UV radiation,
which leads to weaker emission \citep[e.g.,][]{Harada+2018, Harada+2019}.
Indeed, it has been reported that the CH$_3$OH/$^{13}$CO ratio becomes small in the bar-end region of NGC~1068, which can be explained by active massive star formation \citep{Tosaki+2017}.
Based on these results, Cluster~1 is considered to extract information on
various stages of star formation, including moderately dense gas, high-density gas, and PDR, and is consistent with the previous studies discussed above.

One notable point in Cluster~1 is the relatively small SHAP contribution of HCN, which has been widely used as a typical dense-gas tracer.
Since Cluster~1 is located near the bar-end and is associated with active star formation, the small contribution of HCN may appear puzzling at first glance.
However, recent studies have re-examined the conventional view that HCN is a highly dense gas tracer directly related to star formation.
\citet{Stuber+2023} pointed out that N$_2$H$^+$ is a more suitable tracer of dense gas
associated with star formation than HCN.
Furthermore, \citet{Tafalla+2021} suggested that traditional dense-gas tracers, such as HCN and CS, do not necessarily trace highly dense gas itself, and that N$_2$H$^+$ is a more appropriate tracer of dense gas.

Based on these considerations, we speculate that the small contribution of HCN in Cluster~1 indicates that HCN may not be well suited as a highly dense gas tracer directly related to star formation.

    \item Cluster~3: 
Cluster~3 represents the negative scores and corresponds to the arms.  
Since the KPC4 scores of Cluster~3 are negative, molecular lines with negative SHAP values can be interpreted as contributing more strongly to the characteristics of Cluster~3.

The third row of Figure~\ref{KPC4_in_detail} shows the SHAP value of the KPC4 scores for Cluster~3.
In this third row of Figure~\ref{KPC4_in_detail}, CH$_3$OH($J$=2$_K$--1$_K$) shows a strong negative correlation, followed by CO($J$=1--0). CH$_3$OH($J$=2$_K$--1$_K$) is widely used as a shock tracer, and CO ($J$=1--0) traces relatively low density and low temperature gas. Based on this, the Kernel PCA and Kernel SHAP results suggest that Cluster~3 likely represents shocks occurring in low density and low temperature gas.
In fact, a high CH$_3$OH/$^{13}$CO ratio has been reported in this region, and this has been interpreted as evidence that methanol is enhanced by shocks \citep{Tosaki+2017}.
Therefore, the results appear to correctly reflect the physical conditions in this region
and are consistent with previous studies.

    \item Cluster~4: 
Cluster~4 represents the negative scores and corresponds to the CND.  
Since the KPC4 scores of Cluster~4 are negative, molecular lines with negative SHAP values can be interpreted as contributing more strongly to the characteristics of Cluster~4.

The fourth row of Figure~\ref{KPC4_in_detail} shows the SHAP value distribution of the KPC4 scores for Cluster~4.
In this fourth row of Figure~\ref{KPC4_in_detail}, there is no clear correlation between the scores and the SHAP values. 
However, the magnitudes of the SHAP values differ significantly among the molecular lines. 
Thus, the SHAP values in cluster~4 are considered to reflect differences in the contributions of molecular lines to the overall CND.
The red line shows the SHAP value's mean of all molecules in Cluster~4, and the light red area represents the dispersion from the mean. 
Using this as a reference, C$_2$H($N$=1--0) and N$_2$H$^+$($J$=1--0) show the highest SHAP values, followed by CN($N$=1--0). 
In contrast, CH$_3$OH($J$=2$_K$--1$_K$) and HCN($J$=1--0) have the lowest SHAP values, with HC$_3$N($J$=10--9) showing similarly low values. 

Since the KPC2 positive scores are mainly contributed by highly dense gas tracers, we interpreted KPC2 as reflecting highly dense gas information in the CND (Section~\ref{subsubsec:KPC1 and KPC2}). 
In contrast, an examination of the SHAP values in Cluster~4 suggests that this cluster is unlikely to primarily reflect density information. As molecular abundances in the CND can vary due to various chemical processes, we therefore investigate below whether the SHAP values in Cluster~4 reflect variations in molecular abundances.

\begin{figure*}[t]

    \begin{center}
        \includegraphics[width=15cm]{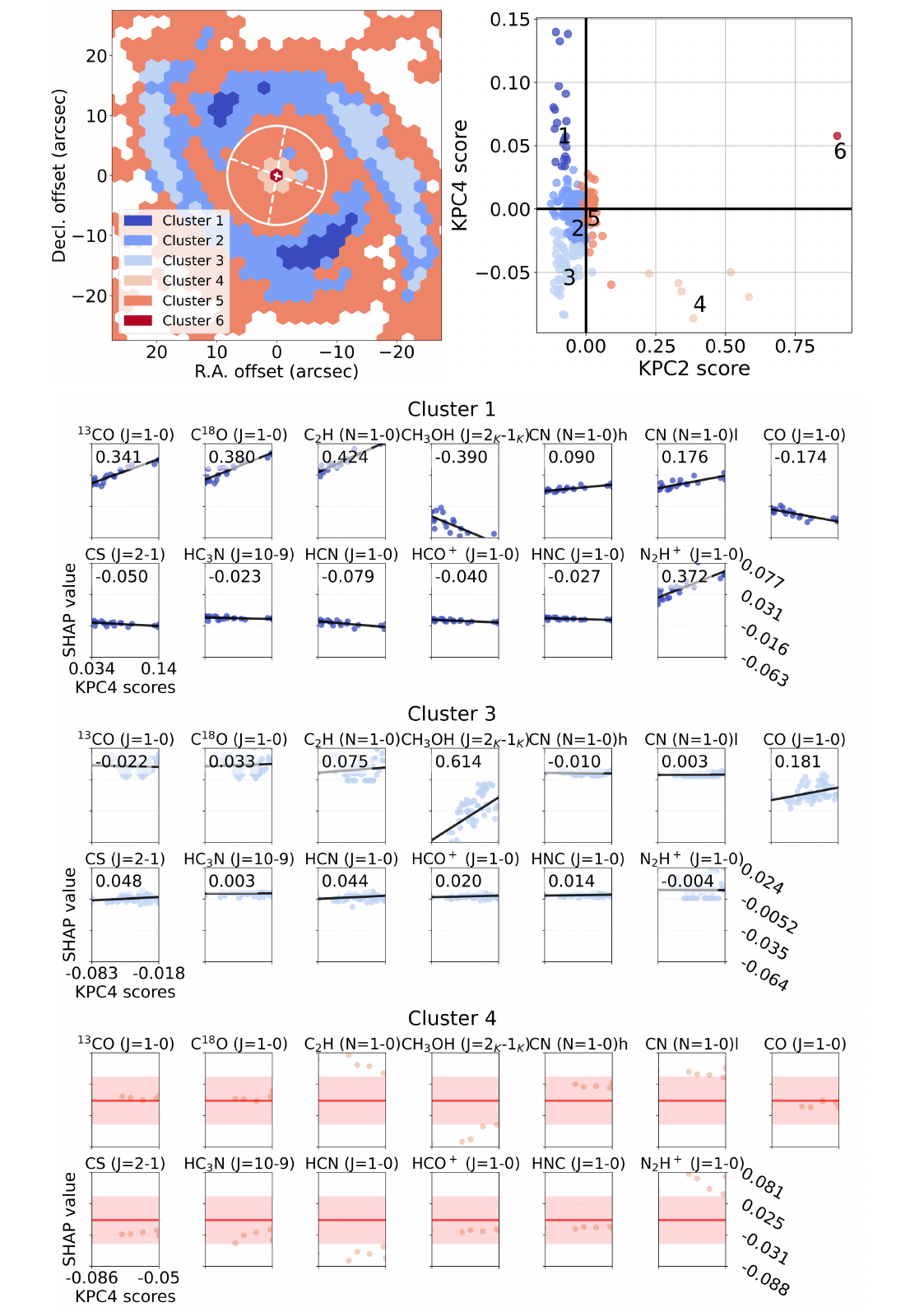}\\
    \end{center}

    \caption{The left panel in the first row shows the cluster map, while the right panel presents the clustering result. The second row illustrates the relationship between the SHAP values and the scores in Cluster~1, and the third row shows the corresponding relationship for Cluster~3. The numbers in the plot represent the correlation coefficients. The fourth row shows the relationship between the SHAP values and the scores in Cluster~4. In each plot, the red line indicates the mean SHAP value, and the light red region represents the 1$\sigma$ range.
    }\label{KPC4_in_detail}
\end{figure*}

In the CND, many processes, such as  X-ray Dominated Region (XDR) conditions, shocks, and high-temperature chemistry, act together in a complex way. Each molecule reflects a different chemical pathway, and we must consider several types of chemistry, including XDR chemistry, CRDR chemistry, shock chemistry, and high-temperature chemistry \citep[e.g.,][]{Meijerink+2005, Izumi+2013, Harada+2013, Huang+2022}.
Molecules show different behaviors depending on the chemical processes in their environment. 

For example, in high-temperature regions, C$_2$H and CN are expected to decrease, and CN can be converted into HCN \citep[e.g., ][]{Harada+2013}. 
In contrast, CN is known to increase in XDR \citep[e.g., ][]{Garcia+2010}. 
Therefore, if the decreasing trend of CN is mitigated compared to that of C$_2$H, it is not contradictory that the SHAP value of CN is smaller than that of C$_2$H.
HCN is also expected to be enhanced in XDR and in several other chemical conditions \citep[e.g., ][]{Kohno+2008, Izumi+2013}. Therefore its low SHAP value is consistent with this previous studies. 
CH$_3$OH is considered a shock tracer, and it has also been reported to decrease in CRDR \citep[e.g.,][]{Watanabe+2003, Aladro+2013}. In the CND of NGC~1068, the abundance of CH$_3$OH has been reported to be enhanced by shock \citep{Huang+2024}. Therefore, if shock chemistry is dominant over CRDR chemistry, the result that the SHAP value of CH$_3$OH is low is not contradictory.
N$_2$H$^+$ can be enhanced at CRDR \citep[e.g.,][]{Aladro+2013, Harada+2015}. 
But it is expected to decrease in XDR \citep[e.g.,][]{Meijerink+2007}. 
The SHAP values of N$_2$H$^+$ suggest that its contribution decreases within the CND. 
The validity of this interpretation is discussed in Section~\ref{subsec:Is N$_2$H$^+$ decreasing in the CND?}.
\end{itemize}

\subsection{Are the abundances of HCN, HCO$^+$, and CS enhanced by the molecular outflow?}\label{subsec: Are the abundances of HCN, HCO$^+$, and CS enhanced by the molecular outflow?}

\begin{table*}[t]
\centering
\caption{KPC3 scores and column densities for each grid}
\setlength{\tabcolsep}{3pt}
\resizebox{\textwidth}{!}{
\begin{tabular}{r r r *{8}{c}}
\hline
$\Delta \mathrm{R.A.}$ & $\Delta \mathrm{Dec.}$ & KPC3 &
$N(^{13}\mathrm{CO})\,[\times10^{15}]$ &
$N(\mathrm{C_2H})\,[\times10^{16}]$ &
$N(\mathrm{CN})\,[\times10^{15}]$ &
$N(\mathrm{CS})\,[\times10^{13}]$ &
$N(\mathrm{HCN})\,[\times10^{13}]$ &
$N(\mathrm{HCO}^+)\,[\times10^{13}]$ &
$N(\mathrm{HNC})\,[\times10^{13}]$ &
$N(\mathrm{N_2H}^+)\,[\times10^{12}]$ \\
\hline
$\mathrm{arcsec}$ & $\mathrm{arcsec}$ &  &
$\mathrm{cm}^{-2}$ & $\mathrm{cm}^{-2}$ & $\mathrm{cm}^{-2}$ &
$\mathrm{cm}^{-2}$ & $\mathrm{cm}^{-2}$ & $\mathrm{cm}^{-2}$ &
$\mathrm{cm}^{-2}$ & $\mathrm{cm}^{-2}$\\
\hline
0.0203 & -3.643 & -0.0866 & 1.29 $\pm$ 0.16 & 1.94 $\pm$ 0.17 & 5.69 $\pm$ 0.51 & 1.72 $\pm$ 0.12 & 1.07 $\pm$ 0.05 & 0.783 $\pm$ 0.059 & 1.30 $\pm$ 0.03 & 0.961 $\pm$ 0.007 \\
3.15 & -1.824 & -0.0789 & 2.68 $\pm$ 0.20 & 0.869 $\pm$ 0.111 & 3.66 $\pm$ 0.43 & 1.61 $\pm$ 0.10 & 1.42 $\pm$ 0.05 & 1.31 $\pm$ 0.08 & 1.21 $\pm$ 0.03 & 2.44 $\pm$ 0.11 \\
-3.11 & -1.819 & -0.173 & 3.62 $\pm$ 0.27 & 4.45 $\pm$ 0.25 & 10.4 $\pm$ 0.80 & 5.48 $\pm$ 0.17 & 3.34 $\pm$ 0.08 & 2.39 $\pm$ 0.09 & 2.70 $\pm$ 0.03 & 2.28 $\pm$ 0.10 \\
-4.15 & 0.00232 & -0.0834 & 2.86 $\pm$ 0.22 & 1.94 $\pm$ 0.21 & 5.16 $\pm$ 0.58 & 2.26 $\pm$ 0.12 & 1.69 $\pm$ 0.07 & 1.24 $\pm$ 0.08 & 1.55 $\pm$ 0.03 & 1.34 $\pm$ 0.09 \\
-3.11 & 1.82 & -0.0842 & 3.68 $\pm$ 0.22 & 1.81 $\pm$ 0.22 & 7.30 $\pm$ 0.69 & 2.26 $\pm$ 0.11 & 1.84 $\pm$ 0.06 & 1.46 $\pm$ 0.09 & 1.61 $\pm$ 0.03 & 2.44 $\pm$ 0.12 \\
0.0154 & 3.64 & -0.105 & 4.34 $\pm$ 0.26 & 4.53 $\pm$ 0.27 & 9.18 $\pm$ 0.67 & 1.66 $\pm$ 0.10 & 1.38 $\pm$ 0.05 & 1.41 $\pm$ 0.08 & 1.67 $\pm$ 0.03 & 1.24 $\pm$ 0.09 \\
2.10 & 3.64 & -0.132 & 3.20 $\pm$ 0.23 & 6.72 $\pm$ 0.30 & 9.12 $\pm$ 0.68 & 1.92 $\pm$ 0.10 & 1.42 $\pm$ 0.04 & 1.47 $\pm$ 0.07 & 1.85 $\pm$ 0.03 & 1.16 $\pm$ 0.10 \\
4.19 & 3.64 & -0.119 & 5.03 $\pm$ 0.22 & 5.86 $\pm$ 0.26 & 7.47 $\pm$ 0.62 & 2.54 $\pm$ 0.11 & 1.56 $\pm$ 0.04 & 1.51 $\pm$ 0.06 & 1.77 $\pm$ 0.03 & 1.11 $\pm$ 0.08 \\
\hline
\end{tabular}
}
\label{tab:col_dens}
\end{table*}

In Section~\ref{subsubsec:KPC3}, we showed that Cluster~2 of KPC3 map traces the molecular outflow.
We also noted that CN, C$_2$H, and HNC are enhanced by UV radiation in the molecular outflow.
On the other hand, Figure~\ref{KPC3_clustering} suggests that HCN and HCO$^+$ may also contribute to the molecular outflow.
Indeed, the abundances of HCN and HCO$^+$ are known to increase or decrease depending on some chemistry \citep[e.g.,][]{Kohno+2005, Narita+2024}.   
In this section, we discuss this possibility in more detail.

If the KPC3 scores in Cluster~2 reflect molecular abundance, the abundance should correlate with the scores. To test this idea, we perform LTE analysis and estimate the column densities in the region of Cluster~2 \citep{Goldsmith+1999}. 
When we calculate the correlation, we remove the grids within 200~pc from the center because these grids are located in the CND.
Since we do not use multiple transitions in this study, we assume that the CO($J$=1--0) line is optically thick and estimate the rotational temperature from its line fitting. 
Here, we do not consider the beam filling factor.
The reason is that the main purpose of this study is to evaluate the correlation between the scores and the column densities.
If the beam filling factor is spatially almost constant, it does not affect the correlation coefficient.
Since the size of one grid is assumed to correspond to one to a few GMCs, the beam filling factor is not expected to vary significantly in space.
In addition, the jet/outflow is suggested to be at a relatively early stage of its evolution, and thus the effect of the outflow on the spatial variation of the filling factor can be neglected \citep[e.g.,][]{May+2017, Michiyama+2022}.
Based on these considerations, beam filling factor does not affect the correlation between the scores and the column densities.
To decide whether a single or a double Gaussian provides a better fit, we use the Akaike Information Criterion (AIC) \citep{Akaike1974}. 
When a double Gaussian is selected, we adopt the average of the two components as the rotational temperature. 
This statistical method is commonly used in astronomical studies\citep[e.g., ][]{Bainbridge+2017, Chen+2022}.

\begin{figure*}[t]

    \begin{center}
        \includegraphics[width=18cm]{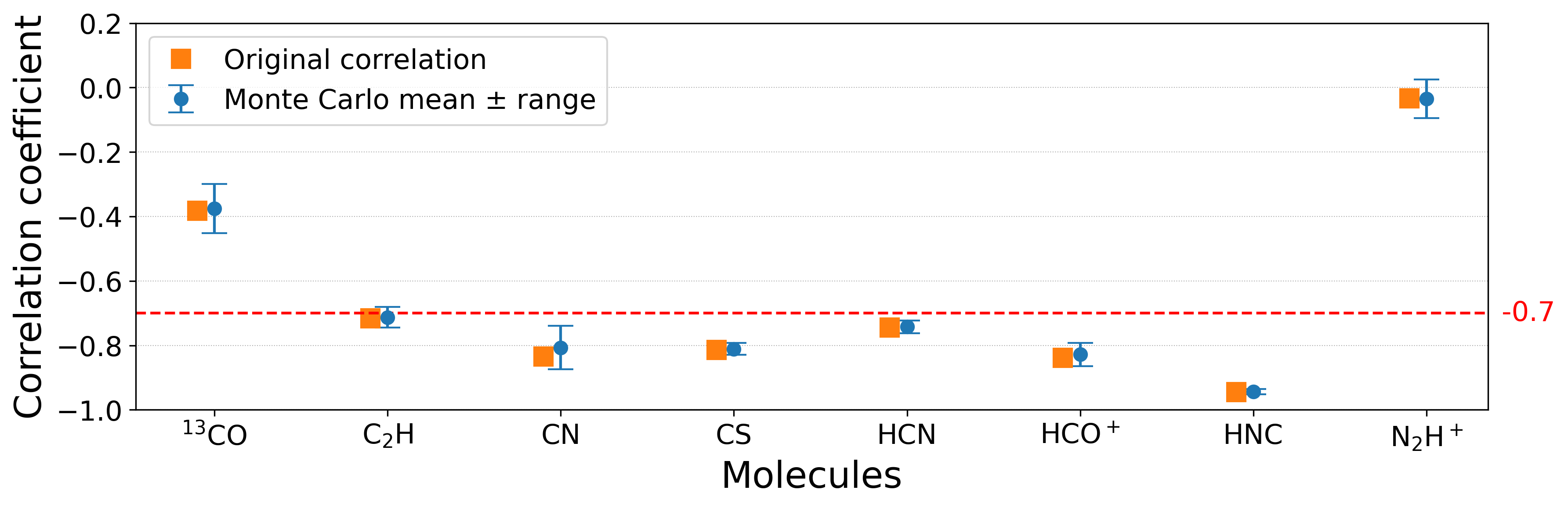}\\
    \end{center}

    \caption{Correlation resutls. The orange squares represent the correlation obtained using the original column densities. The blue circles represent the mean correlation derived from Monte Carlo simulations, and the error bars indicate the $\pm1\sigma$ uncertainty estimated from the Monte Carlo results.}\label{corr_corner}
\end{figure*}

Among the Cluster~2 grids after removing the CND, we use only the molecular lines for which more than half of the grids (eight or more) satisfy S/N~$>$~5. Under these conditions, we compute the column densities of the ALMA Band~3 (12m+7m) lines of $^{13}$CO($J$=1--0), CS($J$=2--1), CN($N$=1--0), C$_2$H($N$=1--0), HCN($J$=1--0), HNC($J$=1--0), HCO$^{+}$($J$=1--0), and N$_2$H$^{+}$($J$=1--0). The results are summarized in Table~\ref{tab:col_dens}.
We calculate their correlation with the KPC3 scores.  
We performed Monte Carlo simulations with 10,000 realizations by randomly perturbing the observed values according to Gaussian distributions, assuming the rms noise in each grid as the $1\sigma$ uncertainty.
The correlation results are shown in Figure~\ref{corr_corner}. 
We find strong negative correlations (correlation coefficients $<-0.7$) between the column densities of CN, C$_2$H, HNC, HCN, HCO$^+$, and CS and the KPC3 scores of Cluster~2.
Because the column density follows $N_{\rm mol} = X_{\rm mol}\,N_{\rm H_2}$ and $N_{\rm H_2}$ is constant for each $N_{\rm mol}$, the correlation with $N_{\rm mol}$ can be regarded as a correlation with abundance. 
As a result, we find that the abundances of CN, C$_2$H, HNC, HCN, HCO$^+$, and CS may be enhanced in the molecular outflow.
This result is not contradictory to the interpretation based on the SHAP values shown in the second row of Figure~\ref{KPC3_clustering}.
In other words, the KPC3 scores in Cluster~2 can be interpreted as capturing variations in molecular abundances associated with the molecular outflow.

We then compare our results with the analysis of \citet{Saito+2022}. 
Their study used [SIII]/[SII] and [CI], which are thought to increase due to UV radiation from the molecular outflow, as labeled data for PCA. 
Here, labeled data refer to observational tracers with a known physical interpretation, which are used as reference indicators in the PCA analysis.
They suggested that the abundances of CN, C$_2$H, and HNC are enhanced. 
In contrast, our analysis does not use any labeled data, yet we find that not only CN, C$_2$H, and HNC, but also HCN, HCO$^+$, and CS may have enhanced abundances. This suggests that the abundances of HCN, HCO$^+$, and CS may be enhanced by chemical processes other than UV radiation (Table~\ref{tab:compare}).

\subsubsection{HCN and CS}\label{subsubsec:HCN and CS}
\citet{Garcia+2014} and references therein concluded that the high intensities of HCN and CS in the molecular outflow region are caused by shock chemistry.
Therefore, the strong correlation between HCN and CS seen in Figure~\ref{corr_corner} is likely to reflect the contribution of shock chemistry in the molecular outflow.

In contrast, studies of the L1157 protostellar outflow showed that HCO$^+$ is not enhanced by shock chemistry \citep[e.g.,][]{Podio+2014}.
Thus, an enhancement of the HCO$^+$ abundance by shock chemistry is unlikely in the molecular outflow of NGC~1068, and the possibility of UV chemistry should be considered.

\subsubsection{HCO$^+$}\label{subsubsec:HCO+}
In general, HCO$^{+}$ is formed primarily through reaction pathways initiated by CO.
However, \citet{Goicoechea+2016} proposed that HCO$^{+}$ is predominantly formed via reaction pathways involving CH$^{+}$ in PDR regions exposed to strong UV radiation and characterized by highly dense gas ($n_{\mathrm{H}} = 10^{6}\ \mathrm{cm}^{-3}$).

According to \citet{Nagy+2013} and \citet{Gonzalez+2023}, CH$^+$-related chemical pathways can be written as follows:
\begin{align}
\mathrm{C^+} + \mathrm{H_2} &\rightarrow \mathrm{CH^+} + \mathrm{H}, \label{eq:CH+}\\
\mathrm{CH^+} + \mathrm{H_2} &\rightarrow \mathrm{CH_2^+} + \mathrm{H}, \\
\mathrm{CH_2^+} + \mathrm{H_2} &\rightarrow \mathrm{CH_3^+} + \mathrm{H}, \\
\mathrm{CH_3^+} + \mathrm{O} &\rightarrow \mathrm{HCO^+} + \mathrm{H_2}.
\end{align}

The molecular outflow region of NGC~1068 is considered to be dense ($n(\mathrm{H_2}) \sim 10^{5\text{--}6}\,\mathrm{cm^{-3}}$) and irradiated by strong UV radiation \citep[e.g.,][]{Garcia+2014, Saito+2022}.
Therefore, it is plausible that HCO$^{+}$ can be formed via reaction pathways involving CH$^{+}$ as suggested by \citet{Goicoechea+2016}.

However, for this mechanism to operate, the presence of CH$^{+}$ itself is required.
Although CH$^{+}$ has been detected in this galaxy, the spatial resolution of the current observations is insufficient to unambiguously determine whether CH$^{+}$ is produced by the molecular outflow \citep[e.g.,][]{Spinoglio+2012}.
We therefore examine the plausibility of CH$^{+}$ formation driven by the molecular outflow.

As shown in Eq.~\eqref{eq:CH+}, the formation of CH$^{+}$ requires a reaction between C$^{+}$ and H$_2$.
This reaction has an activation barrier of approximately 4600~K and thus does not proceed efficiently
under typical molecular cloud conditions \citep[e.g.,][]{Godard+2014}.
However, H$_2$(1--0 S(1)) emission has been detected in this galaxy, and its spatial distribution
is aligned with the direction of the molecular outflow \citep{Davies+1998}.
According to \citet{Davies+1998}, this H$_2$ emission is interpreted as being excited by shocks in molecular clouds within the bar.
Nevertheless, because the directions of the bar and the outflow are aligned, it is also possible
that the H$_2$ emission is excited by UV radiation associated with the outflow.
Indeed, studies of NGC~253 have reported that UV pumping plays a more significant role than shocks
in producing H$_2$ emission \citep[e.g.,][]{Harrison+1998, Rosenberg+2013}.
The upper energy level of the H$_2$(1--0 S(1)) transition corresponds to an excitation temperature
of approximately 7000~K, which is sufficient to overcome the activation barrier and enable the
formation of CH$^{+}$.

Based on these considerations, we suggest that CH$^{+}$ formation is promoted
by UV pumping in the outflow region.
Consequently, HCO$^{+}$ is subsequently formed under conditions of strong UV irradiation and highly dense gas in the outflow region.

\begin{table*}[t]
\centering
\caption{Comparison between \cite{Saito+2022} and this study}

\setlength{\tabcolsep}{3pt}

\resizebox{\textwidth}{!}{%
{\footnotesize
\begin{tabular}{c|c|c|c|c}
\hline
                  & Analysis    & Labeled data       & A possible enhancement              &  Chemistry  \\
\hline
\cite{Saito+2022} & PCA         & [SIII]/[SII], [CI] & CN, C$_2$H, HNC                     &  UV    \\
\hline
This study        & Kernel PCA  & None               & CN, C$_2$H, HNC, HCN, HCO$^+$, CS   &  UV + others\\
\hline
\end{tabular}
} %
} %

\label{tab:compare}
\end{table*}

\subsection{Is N$_2$H$^+$ decreasing in the CND?
}\label{subsec:Is N$_2$H$^+$ decreasing in the CND?}
In Section~\ref{subsubsec:KPC4}, we mentioned that N$_2$H$^+$ may be decreasing in the CND of NGC~1068. 
In this section, we discuss whether this interpretation is reasonable.

First, the formation and destruction of N$_2$H$^+$ are described by the chemical reactions shown below \citep[e.g.,][]{Turner+1995, Molek+2007}.  
\begin{align}
\mathrm{H_3^+} + \mathrm{N_2} &\rightarrow \mathrm{N_2H^+} + \mathrm{H_2}, \\
\mathrm{N_2H^+} + e^- &\rightarrow \mathrm{N_2} + \mathrm{H}, \label{N2H+ decreasing 1}\\
\mathrm{N_2H^+} + e^- &\rightarrow \mathrm{NH} + \mathrm{N}. \label{N2H+ decreasing 2}
\end{align}

From these reactions, we can see that H$_3^+$ is important for the formation, and electrons are important for the destruction.
For example, in CRDR, H$_3^+$ is efficiently produced, which in turn enhances the abundance of N$_2$H$^+$ \citep[e.g.,][]{Harada+2015}.
On the other hand, according to \citet{Pereira-Santaella+2024}, X-ray irradiation destroys H$_2$ which is the parent molecule of H$_3^+$. As a result, the abundance of H$_3^+$ becomes lower.  
In addition, the electron abundance becomes high in XDR, so H$_3^+$ decreases further through recombination.
In addition, Eqs.~\eqref{N2H+ decreasing 1} and \eqref{N2H+ decreasing 2} show that N$_2$H$^{+}$ is destroyed via dissociative recombination with electrons.

The CND of NGC~1068 is thought to be an XDR, and a high electron abundance of $10^{-6}$--$10^{-4}$ has been reported \citep{Usero+2004}.  
Therefore, it is reasonable to speculate that the N$_2$H$^{+}$ abundance may be decreasing in the CND of NGC~1068.
In other words, the high SHAP value of N$_2$H$^{+}$ in Cluster~4 may suggest a possible decrease in the N$_2$H$^{+}$ abundance associated with the XDR environment.

\section{SUMMARY AND CONCLUSIONS}\label{sec:summary}
We combined two independent methods: Kernel PCA, an unsupervised machine-learning technique, and Kernel SHAP, an XAI method.
This combination allows us to extract features from multi-line data more appropriately than conventional PCA.
To compare the performance with conventional PCA, we applied the proposed framework to the NGC~1068 ALMA Band~3 data, for which PCA-based studies have already been conducted.

\begin{enumerate}
  \item We confirmed that the proposed framework can extract characteristic regions and phenomena in the galaxy in more detail than conventional PCA. 
  For this dataset, meaningful physical interpretations are limited to the second principal component in PCA, whereas they can be extended up to the fourth component in this framework.
  \item By comparing the molecular column densities derived from the LTE analysis with the KPC3 scores, we suggest that HCO$^{+}$ is enhanced in the outflow regions due to the effects of UV irradiation and highly dense gas.
  \item These results demonstrate that our framework enables a data-driven characterization of physical and chemical features that were not clearly distinguished in previous studies. 
  Moreover, it offers an efficient approach for interpreting the growing volume of multi-line observational data.
\end{enumerate}

Finally, although we applied our framework to multi-line data in this study, its application can be extended to other types of analyses.
PCA is widely used in multi-object studies and has also been applied to cube data \citep[e.g.,][]{Steiner+2009, Paris+2011, Logan+2020, Donnan+2024}.
Therefore, our framework may also be useful for such studies.

\begin{acknowledgments}
This paper makes use of the following ALMA data: ADS/JAO.ALMA\#2011.0.00061.S, ADS/JAO.ALMA\#2012.1.00657.S, ADS/JAO.ALMA\#2013.1.00060.S, ADS/JAO.ALMA\#2015.1.00960.S, ADS/JAO.ALMA\#2017.1.00586.S, ADS/JAO.ALMA\#2018.1.01506.S, ADS/JAO.ALMA\#2018.1.01684.S, and ADS/JAO.ALMA\#2019.1.00130.S. 
This work was supported by JST SPRING, Grant Number JPMJSP2124.
TTT has been supported by the Japan Society for the Promotion of Science (JSPS) Grants-in-Aid for Scientific Research (21H01128 and 24H00247). 
TTT has also been supported in part by the Collaboration Funding of the Institute of Statistical Mathematics ``Machine-Learning-Based Cosmogony: From Structure Formation to Galaxy Evolution''. 
KAKENHI Grant-in-Aid for Scientific Research(B): Ministry of Education, Culture, Sports, Science and Technology through the Japan Society for the Promotion of Science (JSPS) [24K02995] and CREST: Ministry of Education, Culture, Sports, Science and Technology through the Japan Science and Technology Agency (JST) [JPMJCR2235] under Core Research for Evolutional Science and Technology (CREST).
We used ChatGPT (OpenAI) to assist with English editing and clarity of the manuscript.
ALMA is a partnership of ESO (representing its member states), NSF (USA) and NINS (Japan), together with NRC (Canada), MOST and ASIAA (Taiwan), and KASI (Republic of Korea), in cooperation with the Republic of Chile. The Joint ALMA Observatory is operated by ESO, AUI/NRAO and NAOJ. This research has made use of the NASA/IPAC Extragalactic Database, which is funded by the National Aeronautics and Space Administration and operated by the California Institute of Technology. The National Radio Astronomy Observatory is a facility of the National Science Foundation operated under a cooperative agreement by Associated Universities, Inc. This research has made use of the SIMBAD database, operated at CDS, Strasbourg, France. Data analysis was in part carried out on the Multi-wavelength Data Analysis System operated by the Astronomy Data Center, National Astronomical Observatory of Japan.
\end{acknowledgments}

%

\vspace{5mm}
\facilities{ALMA (Band 3)}


\software{
    ALMA Calibration Pipeline, 
    Astropy \citep{AstropyCollaboration+2013, AstropyCollaboration+2018}, 
    CASA \citep{CASA+2022}, 
    PHANGS-ALMA Pipeline \citep{Leroy+2021},
    NumPy \citep{Harris+2020}, 
    Matplotlib \citep{Hunter+2007}, 
    Scikit-learn \citep{Pedregosa+2011},
    shap \citep{Lundberg+2017}.
}




\bibliography{Okubo25}{}
\bibliographystyle{aasjournal}



\end{document}